\documentclass[fleqn,usenatbib]{mnras}

\usepackage{newtxtext,newtxmath}
\usepackage[T1]{fontenc}
\usepackage{ae,aecompl}
\usepackage{graphicx}	% Including figure files
\usepackage{amsmath}	% Advanced maths commands
\usepackage{subfigure}  % Subfigures
\usepackage{placeins}   % FloatBarrier
\usepackage{tabularx}   %Table with fixed width
\usepackage{mathrsfs}   %for slanted H
\usepackage{dsfont}     %for ganze zahlen Z
\usepackage{rotating}   %for sideways figure
\usepackage{layouts}
\usepackage{enumitem}   %for enumerate functionality
\usepackage{url}

 \newcommand{\vect}[1]{\mathbf{#1}}
\newcommand*\diff{\mathop{}\!\mathrm{d}}

\newcommand{\appropto}{\mathrel{\vcenter{
  \offinterlineskip\halign{\hfil$##$\cr
    \propto\cr\noalign{\kern2pt}\sim\cr\noalign{\kern-2pt}}}}}
 \usepackage[dvipsnames]{xcolor}

\defcitealias{2021MNRAS.500.2645T}{Paper I}
\defcitealias{2019MNRAS.484.3291T}{T19}
\urldef\myurl\url{https://github.com/jan-rybizki/gdr2_completeness/blob/master/tutorials/%5B2%5DCompleteness%20tutorial_gdr2_light.ipynb}
\urldef\myurlrvs\url{https://github.com/jan-rybizki/gdr2_completeness/blob/master/tutorials/%5B5%5D%20RVS_selection_function.ipynb}

\title[The Galactic bar in angle space]{Identifying resonances of the Galactic bar in \emph{Gaia} DR2:\\II. Clues from angle space}

\author[W.~H.~Trick]{Wilma H.~Trick$^{1}$\thanks{E-mail: trick@mpa-garching.mpg.de}
\\
$^{1}$Max-Planck-Insitut f\"ur Astrophysik, Karl-Schwarzschild-Str. 1, D-85748 Garching b. M\"unchen, Germany
}

\date{Accepted XXX. Received YYY; in original form ZZZ}

\pubyear{2019}

\begin{document}
\label{firstpage}
\pagerange{\pageref{firstpage}--\pageref{lastpage}}
\maketitle

% Abstract of the paper
\begin{abstract}
The Milky Way disk exhibits intricate orbit substructure of still-debated dynamical origin. The angle variables $(\theta_\phi,\theta_R)$---which are conjugates to the actions $(L_z,J_R)$, and describe a star's location along its orbit---are a powerful diagnostic to identify $l$:$m$ resonances via the orbit shape relation $\Delta \theta_R / \Delta \theta_\phi = -m/l$. In the past, angle signatures have been hidden by survey selection effects (SEs). Using test particle simulations of a barred galaxy, we demonstrate that \emph{Gaia} should allow us to identify the Galactic bar's Outer Lindblad Resonance ($l=+1, m=2$, OLR) in angle space. We investigate strategies to overcome SEs. In the angle data of the \emph{Gaia} DR2 RVS sample, we independently identify four candidates for the OLR and therefore for the pattern speed $\Omega_\text{bar}$. The strongest candidate, $\Omega_\text{bar}\sim1.4\Omega_0$, positions the OLR above the `Sirius' moving group, agrees with measurements from the Galactic center, and might be supported by higher-order resonances around the `Hercules/Horn'. But it misses the classic orbit orientation flip, as discussed in the companion study on actions. The candidate $\Omega_\text{bar}\sim1.2\Omega_0$ was also suggested by the action-based study, has the OLR at the `Hat', is consistent with \emph{slow bar} models, but still affected by SEs. Weaker candidates are $\Omega_\text{bar}=1.6\Omega_0$ and $1.74\Omega_0$. In addition, we show that the stellar angles do not support the `Hercules/Horn' being created by the OLR of a \emph{fast bar}. We conclude that---to resolve if `Sirius' or `Hat' is related to the bar's OLR---more complex dynamical explanations and more extended data with well-behaved SEs are required.
\end{abstract}

% Select between one and six entries from the list of approved keywords.
% Don't make up new ones.
\begin{keywords}
Galaxy: disc -- Galaxy: kinematics and dynamics
\end{keywords}

\section{Introduction}

Two puzzles have engaged Galactic astronomers for over two decades, and an unambiguous answer would be a milestone in understanding the mechanisms that have shaped our Milky Way (MW):
\begin{enumerate}[leftmargin=*,topsep=1ex,itemsep=1ex,label=(\arabic*)]
    \item What is the value of the Galactic bar's pattern speed, $\Omega_\text{bar}$?
    \item Which features observed in the local stellar phase-space are due to bar resonances?
\end{enumerate}
These two aspects of the Galactic structure are coupled via the resonance condition
    \begin{equation}
        m \cdot \left(\Omega_{\phi,\text{true}} - \Omega_\text{bar} \right) + l \cdot \Omega_{R,\text{true}} = 0, \label{eq:general_res_cond}
    \end{equation}
(where $\Omega_{\phi,\text{true}}$ and $\Omega_{R,\text{true}}$ are the primary azimuthal and radial frequency of a stellar orbit, and $l,m \in \mathds{Z}$), and via the Galaxy's gravitational potential $\Phi(\vect{x})$, in particular:
\begin{enumerate}[leftmargin=*,topsep=1ex,label=(\roman*)]
    \item the rotation curve $v_\text{circ}(R\mid\Phi)$ which sets the orbital frequencies and therefore the location of the bar resonances in the disk,
    \item the mass and shape of the bar which set the strength and number of the local resonance features (e.g. \citealt{2018MNRAS.477.3945H,2019AA...626A..41M}), 
    \item additional mechanisms that might affect and obscure the bar resonances (e.g. spiral arms \citep{2003AJ....125..785Q,2010ApJ...722..112M,2011MNRAS.417..762Q,2016MNRAS.461.3835M,2019MNRAS.482.1983F,2019MNRAS.490.1026H,2020MNRAS.491.2162P}, satellite interaction \citep{2019MNRAS.489.4962K,2019MNRAS.485.3134L}, or bar deceleration \citep{2021MNRAS.500.4710C}).
 \end{enumerate}
 
The above questions (1) and (2) have been studied in the past together or independently, using different approaches and data sets.

Recent studies measured bar pattern speeds around $\Omega_\text{bar} \sim40~\text{km/s/kpc}$ directly, e.g. from \emph{Gaia} proper motions \citep{2016A&A...595A...1G} observed in the Galactic center \citep{2019MNRAS.488.4552S,2019MNRAS.489.3519C,2019MNRAS.490.4740B} using the \citet{1984ApJ...282L...5T} method, from made-to-measure modeling of red clump stars \citep{2017MNRAS.465.1621P}, or interpreting observed gas flows \citep{2015MNRAS.454.1818S}.

The `Hercules' stream is a prominent feature in the stellar velocities within $\sim 200~\text{pc}$ from the Sun at $V\sim-50~\text{km/s}$. It has been noted to resemble the expected signature of the 1:2 Outer Lindblad Resonance ($l=+1,m=2$, OLR for short) of a \emph{short fast bar} (e.g. \citealt{2000AJ....119..800D,2001AA...373..511F,2014AA...563A..60A,2017MNRAS.466L.113M}), of the Co-rotation Resonance ($l=0$, CR) of a \emph{long slow bar} (e.g. \citealt{2017ApJ...840L...2P,2019A&A...626A..41M,2020MNRAS.495..895B,2020ApJ...890..117D}), or of the outer 1:4 Lindblad resonance ($l=+1,m=4$) of a \emph{slightly faster slow bar} (\citealt{2018MNRAS.477.3945H}).

The precision and accuracy of the stellar velocities measured by the \emph{Gaia} satellite's second data release (DR2, \citealt{2018A&A...616A...1G,2019A&A...622A.205K}) have made local moving groups visible in unprecedented detail \citep{2018AA...616A..11G}. Several authors have noted that the `Hat' moving group at $V\sim40~\text{km/s}$ could look like an OLR also---in this case caused by a \emph{slow bar} \citep{2019A&A...626A..41M,2019MNRAS.490.1026H,2021MNRAS.500.2645T}.

The large spatial extent of \emph{Gaia} DR2's 6D stellar phase-space data has spiked the interest in studying the features in the disk's orbit distribution and bar resonances in the disk beyond the immediate Solar neighbourhood. These orbit features are prevalent over a few kpc from the Sun and project locally into the classic moving groups (\citealt{2019MNRAS.484.3291T}, \citetalias{2019MNRAS.484.3291T} hereafter). They have been studied in the $(R,v_\phi)$ plane \citep{2018MNRAS.479L.108K}, where the OLR creates elongated arches (e.g., \citealt{2019MNRAS.488.3324F}), the $(L_z,\phi)$ plane, where the sloping of the orbit structures could be informative about the resonances (e.g., \citealt{2019A&A...632A.107M,2020ApJ...890..117D,2019MNRAS.490.5414F,2021MNRAS.500.4710C}), and action space $(L_z,J_R)$ \citepalias{2019MNRAS.484.3291T}, where resonances create high-$J_R$ ridges (e.g., \citealt{2010MNRAS.409..145S,2019MNRAS.484.3154S,2019AA...626A..41M,2019MNRAS.490.1026H,2021MNRAS.500.2645T}).

The discrepancy between the \emph{slow bar} model from the Galactic center modeling and the \emph{fast bar} from the `Hercules' stream modeling is a classic conundrum of Galactic dynamics. Recently, a new inconsistency has emerged: The newest Galactic center measurements of the bar pattern speed do not position their OLR close to any plausible feature in the local kinematics. \citet{2019AA...626A..41M} mentions that the \emph{slow bar} by \citet{2017MNRAS.465.1621P} and \citet{2017ApJ...840L...2P} requires modifications to the MW's assumed rotation curve or Solar motion (e.g. \citealt{2010MNRAS.403.1829S}) to consolidate the OLR with the `Hat' feature (at $V\sim40~\text{km/s}$). \citet{2019MNRAS.490.1026H} also noticed that for the \emph{slow bar} pattern speed, the resonant ridges fall in the wrong phase-space locations to agree with the \emph{Gaia} data. In \citet{2021MNRAS.500.2645T}, we showed that by combining the \emph{slightly faster slow bar} $\Omega_\text{bar}$ by \citet{2019MNRAS.488.4552S} and \citet{2019MNRAS.490.4740B} with recent models for the MW's rotation curve (\citealt{2015ApJS..216...29B,2019ApJ...871..120E}), the OLR falls close to the \texttt{Sirius} ridges (with the associated `Sirius' moving group being at $(U,V) = (10,3)~\text{km/s}$), a region in phase-space that does not agree with the naive expectation for an OLR signature.

Several intriguing proposals could resolve this new contradiction. (A) \citet{2021MNRAS.500.4710C} have studied a decelerating bar model---as opposed to the often used bar models with a constant pattern speed---which creates signatures that could indeed be closer to the observed ones around the `Sirius' moving group with the current OLR at the `Hat'. (B) Galaxy simulations suggest that in more complex environments, the most reliable signature of the OLR is the high-$J_R$ ridge it creates \citep{2020MNRAS.494.5936F,2021MNRAS.508..728K}, as spiral arms might obscure the idealized $U \sim v_R$ pattern of the OLR \citep{2019MNRAS.482.1983F}.  (C) Spiral arm models can create ridges just as strong as those observed in the \emph{Gaia} data (e.g., \citealt{2004MNRAS.350..627D,2009ApJ...700L..78A,2019MNRAS.484.3154S,2019MNRAS.490.1026H}). Work by \citet{2018MNRAS.480.3132Q} and \citet{2020A&A...634L...8K} associates the main features we see in the local data not with the bar resonances, but with the MW's spiral arms as identified by \citet{2014ApJ...783..130R} and \citet{2016SciA....2E0878X}: the `Hat' to the Perseus arm (see also \citealt{2017MNRAS.467L..21H}), `Sirius' to the Local Arm, the `Horn' (at $(U,V)\sim (50,-20)~\text{km/s}$) to the Sagittarius arm, and `Hercules' to the Scutum arm. (D) Our knowledge of the MW's gravitational potential and the Sun's motion might be the culprit \citep{2017MNRAS.465.1443M,2019AA...626A..41M}.

\begin{figure*}
    \centering
    \subfigure[The azimuthal angle $\theta_\phi$ is the azimuth of the guiding center. \label{fig:epicycle_approx_theta_phi}]{
        \includegraphics[width=0.45\textwidth]{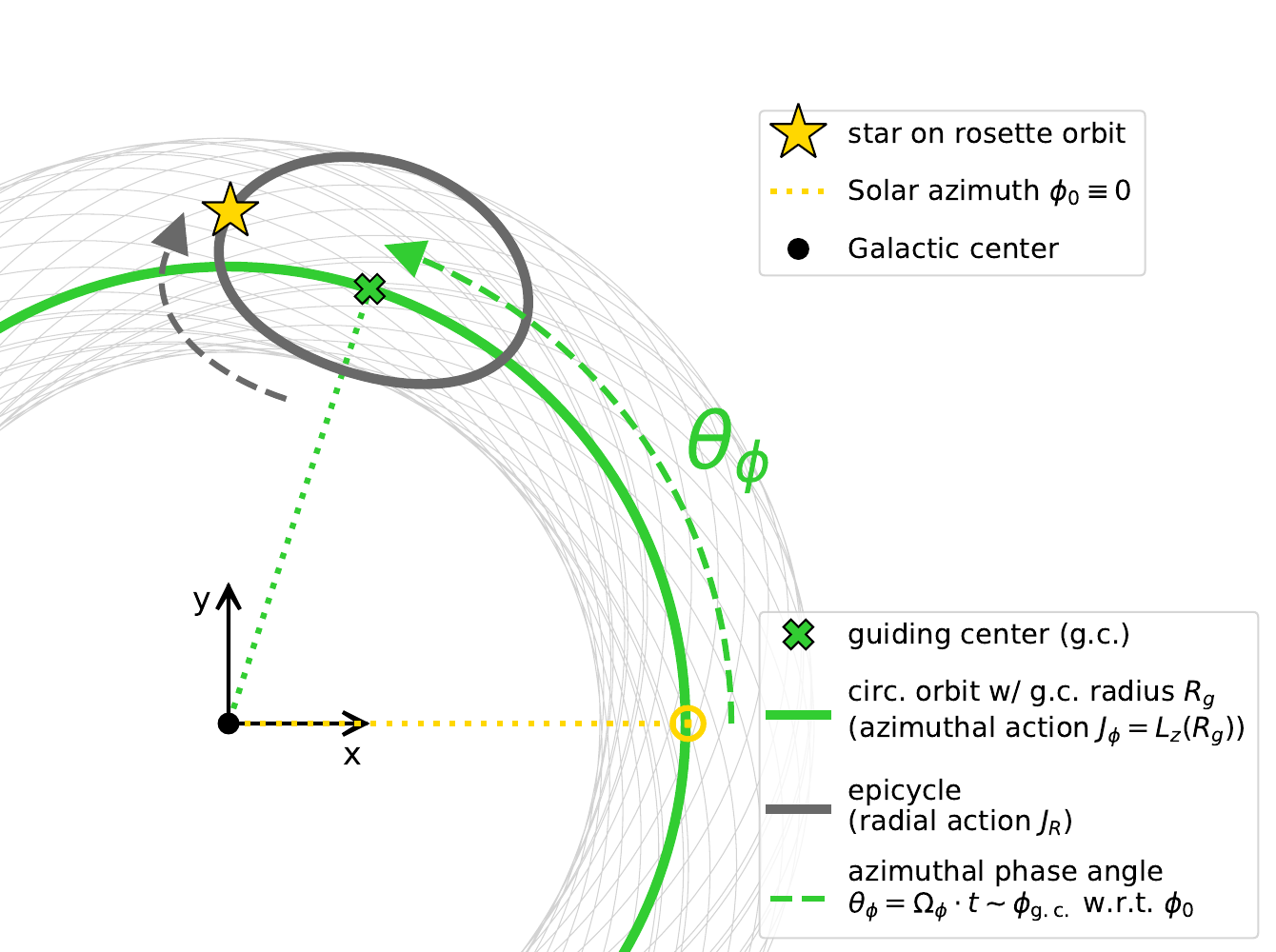}
    }\hfill
    \subfigure[The radial angle $\theta_R$ describes the radial phase between peri- and apocenter. \label{fig:epicycle_approx_theta_R}]{
        \includegraphics[width=0.45\textwidth]{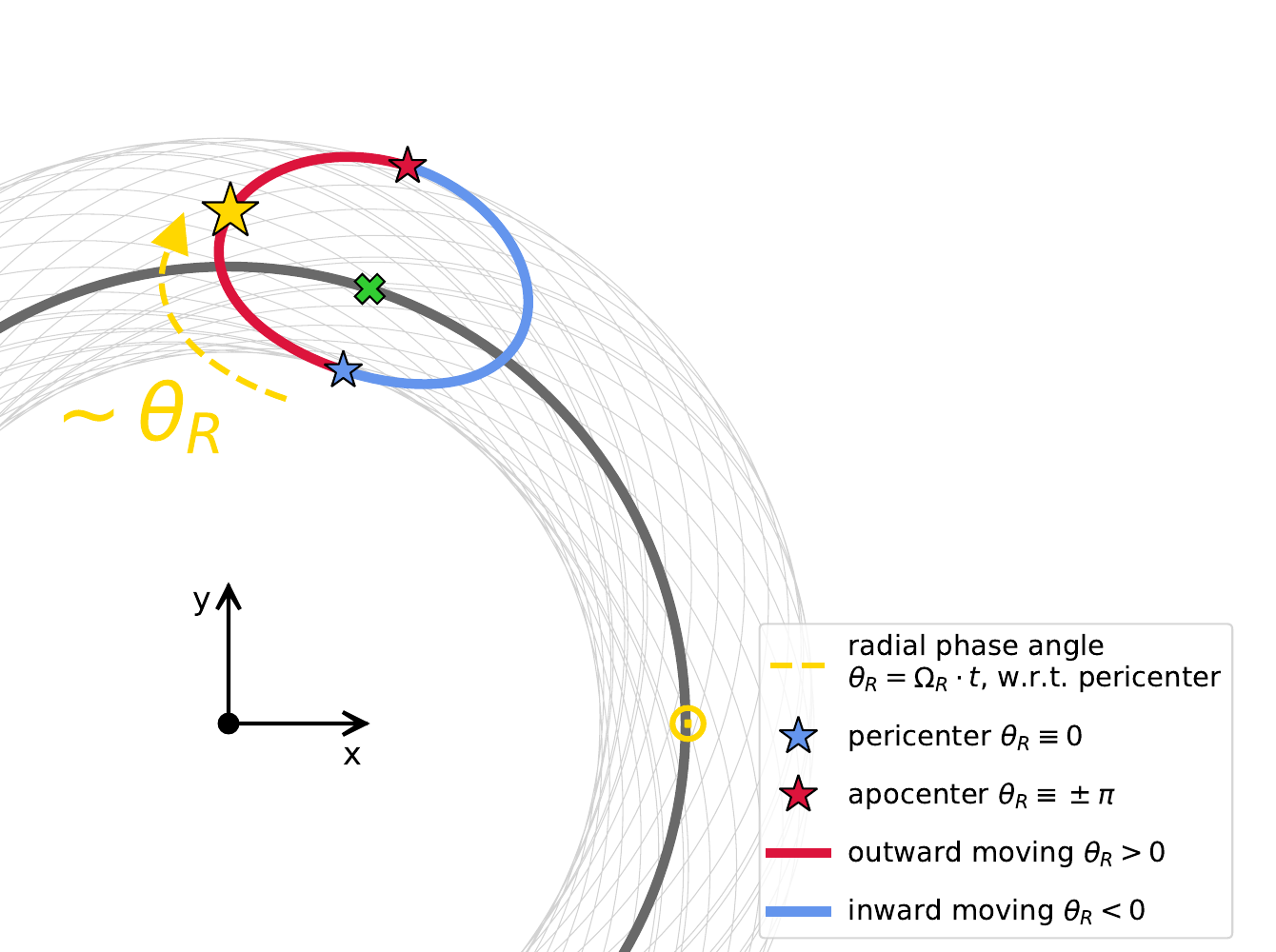}
    }
    \caption{Illustration of how to interpret the in-plane phase angles $\theta_\phi$ and $\theta_R$ associated with the actions $J_\phi\equiv L_z$ and $J_R$ in the epicycle approximation for a disk orbit in an axisymmetric galaxy disk.}
    \label{fig:angles_in_epicycle_approximation_explained}
\end{figure*}

In summary, the complexity and degeneracy of all the mechanisms possibly at play in the Galactic disk call for (a) a deeper understanding of the effect different mechanisms can have alone and together, and (b) straightforward, discriminating diagnostics that help to determine which mechanism causes which orbital feature in the \emph{Gaia} data. Extending our investigations in \citet{2021MNRAS.500.2645T} (\citetalias{2021MNRAS.500.2645T}, hereafter), we focus in this work therefore on (a) bar resonances in a basic MW model (a galactic disk, perturbed by a boxy bar with a constant pattern speed and without spiral arms), and study (b) the signatures caused in the space of \emph{orbital phase angles}. The radial phase $\theta_R$ and the azimuthal phase $\theta_\phi$ describe the location of a star on its orbit. They are canonical conjugates to the orbital actions $J_R$ and $J_\phi=L_z$, respectively. Their physical interpretation is illustrated in Figure \ref{fig:angles_in_epicycle_approximation_explained}.

The diagnostic power of phase-angles has previously been noted in the literature. \citet{2010MNRAS.409..145S} showed that the `Hyades' moving group (around $(U,V) = (-33, -16)~\text{km/s}$) is consistent with being created by an inner ($l=-1$) Lindblad resonance, based on the idea that angle-space can be used to identify the exact $(l,m)$ of a resonant feature. Follow-up studies by \citet{2011MNRAS.418.1565M} and \citet{2011MNRAS.418.2459H} agreed with \citet{2010MNRAS.409..145S} that \emph{action}-space indeed supports the resonant origin of the `Hyades', which is also in concordance with age-abundance measurements \citep{2007A&A...461..957F,2011MNRAS.415.1138P}. However, they pointed out that for the $d<200~\text{pc}$ survey volume of the pre-\emph{Gaia} era, selection effects in \emph{angle}-space are too severe to unambiguously determine the $(l,m)$ of the resonance at the `Hyades'. They all agreed that a larger coverage of data in the Galactic disk might help with disentangling physical signatures from selection biases in angle space. In Section \ref{sec:discussion_Hyades_papers}, we will discuss the work by \citet{2010MNRAS.409..145S} and \citet{2011MNRAS.418.1565M} in more detail, in the context of comparing to our results.

The exquisite data quality of \emph{Gaia} DR2 motivated \citet{2019MNRAS.484.3154S} and \citet{2019MNRAS.490.1026H} to revisit the Solar neighbourhood $d<200~\text{pc}$ in angle space, making the intricate disk substructure beyond the `Hyades' visible. As pointed out in fig. 3 in \citet{2019MNRAS.490.1026H} (and as will be discussed in Section \ref{sec:selecting_resonant_stars}), this provides an alternative way to show the well-known moving groups.

This paper is structured as follows. In Section \ref{sec:theory}, we revisit resonant orbits in action-angle space. Inspired by the study of \citet{2010MNRAS.409..145S}, we present in Sections \ref{sec:simulation_description}-\ref{sec:selecting_resonant_stars} a method based on angle-space to identify the bar OLR directly from the \emph{Gaia} data. Cautioned by the study of \citet{2011MNRAS.418.1565M}, we develop a strategy to disentangle physical signatures from selection effects in Sections \ref{sec:selection_effects_with_distance}-\ref{sec:selection_effects_with_longitude}, even though we do not know \emph{Gaia}'s detailed completeness function. In Section \ref{sec:data_description}-\ref{sec:OLR_candidates}, we search for the bar's OLR in the in-plane angle space of the \emph{Gaia} DR2 radial velocity sample (RVS) \citep{2019A&A...622A.205K} out to $d=3~\text{kpc}$, and present our favoured candidates for the corresponding bar pattern speed $\Omega_\text{bar}$. In the discussion, Section \ref{sec:discussion}, we compare these candidates with the OLR candidates found from the method based on action-$v_R$-space in \citetalias{2021MNRAS.500.2645T}, present a curious coincidence related to higher-order resonances, and discuss previous work. For detailed discussions of the bar pattern speed and OLR candidates in the literature, we refer the reader to \citetalias{2021MNRAS.500.2645T}. We summarize in Section \ref{sec:summary}.

\section{Resonant orbits in angle space} \label{sec:theory}

\subsection{The coordinates: Axisymmetric action-angle estimates} \label{sec:axisym_action_angles}

A star on a near-circular in-plane orbit in the Galactic disk follows a rosette-like path. Its motion can be decomposed into a circular orbit with angular momentum $L_z$ and a radial epicycle motion (Figure \ref{fig:angles_in_epicycle_approximation_explained}). The star moves along the epicycle around the guiding center with the radial frequency $\Omega_R$; the guiding center moves around the Galactic center with the tangential frequency $\Omega_\phi$. The size of the circular orbit is related to the azimuthal action 
\begin{equation}
J_\phi \equiv L_z \equiv R_g \times v_\text{circ}(R_g) = R \cdot v_\phi,
\end{equation}
where $R_g$ is the guiding center radius, $v_\text{circ}^2 = \left. R \times \partial \Phi / \partial R \right|_{z=0}$, $(R,z,\phi)$ Galactocentric cylindrical coordinates, and $(v_R,v_\phi,v_z)$ the velocities in the corresponding directions. The size of the epicycle is related to the radial action
\begin{equation}
J_R \equiv \frac{1}{\pi} \int_{u_\text{min}}^{u_\text{max}} p_u \diff u \in [0,\infty).
\end{equation}
In the prolate confocal coordinate system $(u,v,\phi)$, $u$ is constant on ellipses in the $(R,z)$ plane. In the plane $(z=0)$, $u$ is therefore a radius-like coordinate \citep[\S 3.5.3]{2008gady.book.....B}. In an axisymmetric gravitational galaxy potential, these two actions are conserved canonical momenta whose conjugate positions are the phase angles $\theta_\phi$ and $\theta_R$. Action-angles have convenient properties:

\begin{eqnarray}
    \dot{J}_i = -\frac{\partial \mathscr{H}_\text{axi}}{\partial \theta_i} =  0 &\Longrightarrow& J_i(t) = \text{const.},  \label{eq:action_def}\\
    \dot{\theta}_i =\frac{\partial \mathscr{H}_\text{axi}}{\partial J_i} =  \Omega_{i,\text{axi}}(\vect{J}) &\Longrightarrow& \theta_i(t) = \Omega_{i,\text{axi}} \cdot t + \theta_{i,0}, \label{eq:angle_def}
\end{eqnarray}

with $i\in\{R,\phi,z\}$, $\vect{J} = (J_R,J_\phi,J_z)$, and where $\mathscr{H}_\text{axi}$ is the Hamiltonian of the axisymmetric system \citep[\S 3.5]{2008gady.book.....B}. $(\vect{\theta},\vect{J})$ describes the same full 6D phase-space as the usual position-velocity space $(\vect{x},\vect{v})$. The angles increase linearly, $\vect{x}$ and $\vect{v}$ are periodic in the angles, and in this work, we define the angles to be $\theta_i \in [-\pi,\pi]$.. The angle $\theta_R$ can be interpreted as the current radial phase position of the star along the epicycle between peri- and apocenter, with respect to the pericenter at $\theta_R \equiv 0$ (Figure \ref{fig:epicycle_approx_theta_R}). The angle $\theta_\phi$ is the current polar azimuth $\phi$ of the guiding center with respect to the azimuth of the Sun today, i.e. $\phi_0 \equiv 0$ (Figure \ref{fig:epicycle_approx_theta_phi}).\footnote{Note that this straightforward interpretation of actions and angles based on the epicycle approximation is only valid for near-circular disk orbits.} A nice explanation of the angles can also be found in \citet[\S 2]{2011MNRAS.418.1565M}.

To calculate these actions, angles, and frequencies, one requires measurements of a star's current position $\vect{x}$ and velocity $\vect{v}$ (e.g. by \citealt{2016A&A...595A...1G}), as well as an assumption for the axisymmetric gravitational galaxy potential in which the star moves (e.g. the \texttt{MWPotential2014} by \citealt{2015ApJS..216...29B}), and an efficient estimation algorithm (e.g. the St\"{a}ckel fudge by \citealt{2012MNRAS.426.1324B,2013ApJ...779..115B}).

Naturally, the Galaxy is not axisymmetric, as it has---among other perturbing mechanisms---also a strong central bar which rotates with a pattern speed $\Omega_\text{bar}$. In this case, actions and angles calculated in the axisymmetric background potential at each point in time still describe coordinates for the stars as valid as positions and velocities. The actions  $(J_R,L_z)$ are, however, not fully conserved anymore (Equation \ref{eq:action_def}), and the angles $(\theta_R,\theta_\phi)$ do not follow the exact linear relation in Equation \eqref{eq:angle_def}. As shown in \citet{2021MNRAS.500.2645T}, these axisymmetric action estimates are still very informative about the perturbed system: In particular the \emph{axisymmetric resonance lines (ARLs)}, i.e. the lines in axisymmetric action space along which the resonance condition is satisfied,
\begin{equation}
    m \cdot (\Omega_{\phi,\text{axi}} - \Omega_\text{bar}) + l \cdot \Omega_{R,\text{axi}}  = 0 \hspace{0.5cm}\text{(ARL)}\label{eq:axisym_res_cond}
\end{equation}
(with $l,m \in \mathds{Z}$, and $J_z =0$), prove to be helpful diagnostic tools.

In this work, we distinguish between two sets of orbital frequencies:
\begin{itemize}[leftmargin=*,topsep=1ex,itemsep=1ex]
    \item $\Omega_{R,\text{true}}$ and $\Omega_{\phi,\text{true}}$ are the real radial and azimuthal orbital frequencies in a barred system. They are determined as the primary frequencies from a Fourier transform of the actual time-evolved orbit in the barred potential (see Section \ref{sec:simulation_description} and Figure \ref{fig:slow_angle_frequencies}).
    \item $\Omega_{R,\text{axi}}$ and $\Omega_{\phi,\text{axi}}$ are the axisymmetric frequency estimates. They are calculated instantaneously in the axisymmetric background potential and are associated with the axisymmetric action-angle estimates in Equation \eqref{eq:angle_def}.
\end{itemize}
These frequencies can differ substantially from each other.

\begin{figure*}
    \centering
    \includegraphics[width=\textwidth]{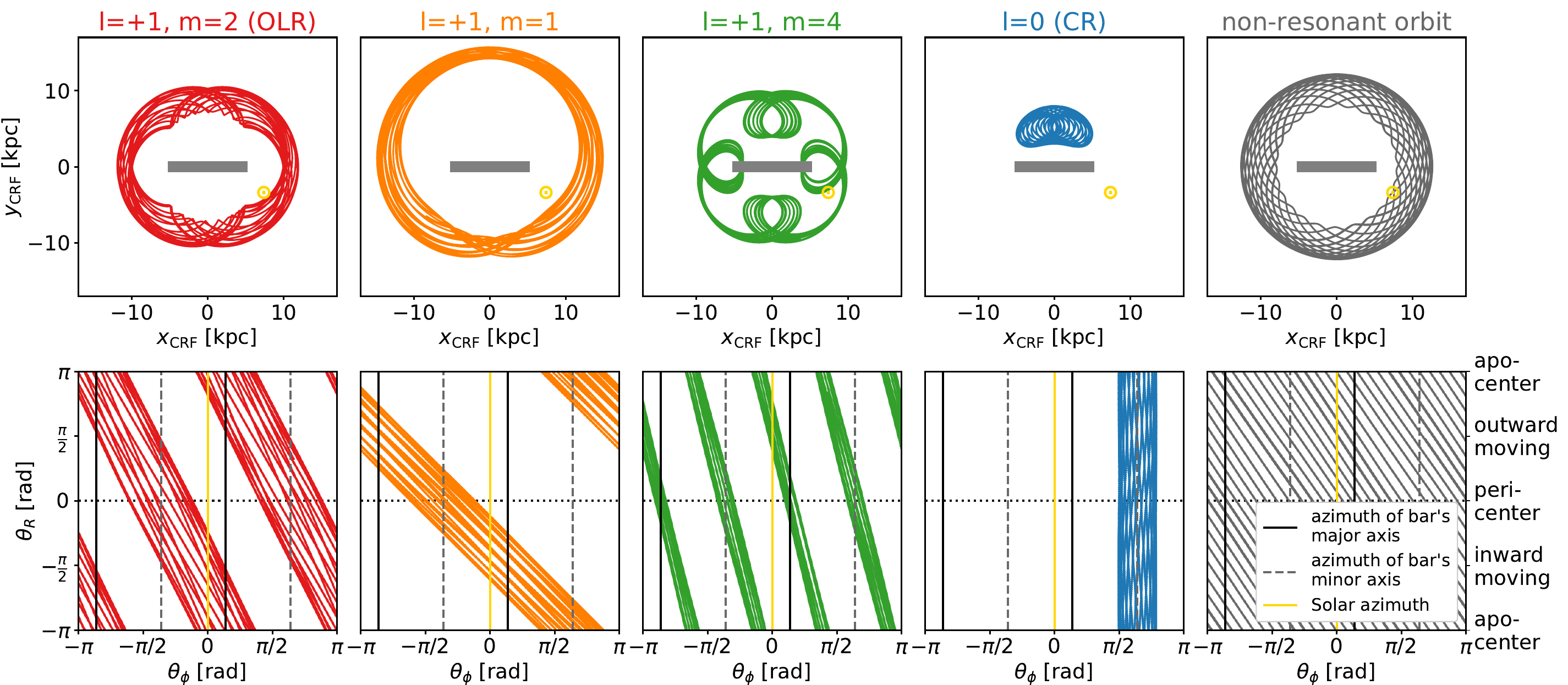}
    \caption{Illustration of five example disk orbits in the \texttt{Fiducial\_40} simulation, trapped at four different $l$:$m$ bar resonances, and one non-resonant orbit. \emph{Upper panels:} Time evolution in the Galactic plane $(x,y)$ in the frame co-rotating (CRF) with the bar. \emph{Lower panels:} The $(\theta_\phi,\theta_R)$ angle space, now assuming the orbit was captured at a single snapshot in time and each dot along the orbit was an individual star. The peri- and apocenters of resonant orbits are `trapped' within a limited azimuthal range with respect to the bar. The distribution in the corresponding angle planes follows $\Delta \theta_R / \Delta \theta_\phi \sim -m/l$ due to the intrinsic orbit shape.}
    \label{fig:example_trapped_orbits}
\end{figure*}

\subsection{Background: Orbits in resonance with the bar}  \label{sec:background_slow_angle_theory}

The behaviour in action-angle space of orbits in bar potentials is well studied in the literature. Introductions based on perturbation theory can be found in \citet{2017MNRAS.471.4314M} and \citet{2021MNRAS.500.4710C}, and a more rigorous derivation in \citet{2018MNRAS.474.2706B}. In \citetalias{2021MNRAS.500.2645T}, we illustrated the action-space behaviour by means of numerically integrated orbits. Here, we summarize the most important concepts, which are also illustrated for one family of OLR orbits in Figure \ref{fig:slow_angle_frequencies} in Appendix \ref{app:slow_angle_frequencies}.

\emph{Rosette orbits} in axisymmetric galaxy potentials have peri- and apocenters at any possible azimuth $\phi \in [-\pi,\pi]$. Over time, such an orbit will fill the whole $(2\pi)^2$ angle plane. The same is true for non-resonant orbits in barred potentials. This orbit characteristic is called \emph{circulation}.

\emph{Resonant orbits}, whose orbital frequencies $\Omega_{i,\text{true}}$ are exactly commensurate with the bar's pattern speed by the ratio $l$:$m$ according to the resonance condition Equation \eqref{eq:general_res_cond}, behave differently. We illustrate this using the axisymmetric action-angle-frequency estimates. The resonance condition for the $\Omega_{i,\text{axi}}$ (Equation \ref{eq:axisym_res_cond}) is not exactly satisfied, but becomes very small,
\begin{equation}
\Omega_\text{slow} \equiv m \cdot (\Omega_{\phi,\text{axi}} - \Omega_\text{bar}) + l \cdot \Omega_{R,\text{axi}}  \sim 0 \label{eq:def_slow_frequency}
\end{equation}
(see, e.g., eq. (11) and (12) in \citealt{2021MNRAS.500.4710C}). By time-integrating Equation \eqref{eq:def_slow_frequency}---using $\phi_\text{bar}(t) \equiv \Omega_\text{bar}t+\phi_{\text{bar},0}$, and Equation \eqref{eq:angle_def}---we get
\begin{equation}
    \theta_\text{slow}(t) \equiv m \cdot \left(\theta_\phi(t) - \phi_\text{bar}(t)\right) + l \cdot \theta_R(t) \sim \text{const.}, \label{eq:slow_angle_def}
\end{equation}
i.e., an axisymmetric estimate for an angle coordinate that evolves very slowly for an orbit close to the $l$:$m$ resonance (e.g., \citealt{1994ApJ...420..597W}). The angles $\theta_R$ and $\theta_\phi$ still evolve fast in time, and the radial phase is often called the \emph{fast angle} of the resonant motion,
\begin{equation}
    \theta_\text{fast} \equiv \theta_R.
\end{equation}

\emph{A family of resonant orbits} trapped at a given $l$:$m$ resonance consists of (i) one closed periodic parent orbit and (ii) librating orbits. The orbits in the same resonant orbit family have the same Jacobi energy
\begin{equation}
    E_\text{Jacobi} = E - \Omega_\text{bar}\cdot L_z = \text{const.},
\end{equation}
where $E\equiv(v_R^2+v_\phi^2)/2+\Phi(R,\phi)$. $E_\text{Jacobi}$ is a conserved quantity in a triaxial, rotating system \citep[eq. 3.112]{2008gady.book.....B}. From this follows directly---using the resonance condition in Equation \eqref{eq:axisym_res_cond}, and $E_i \sim J_i\Omega_{i,\text{axi}}$ as the energy stored in the motion in the $i$-th coordinate direction---that they also have a similar values of the \emph{fast action}
\begin{equation}
    J_\text{fast} \equiv J_R - \frac{l}{m} L_z \sim \text{const.}
\end{equation}
\citep{1979MNRAS.187..101L,1994MNRAS.268.1041K,1994ApJ...420..597W}. All these orbits oscillate in the $(L_z,J_R)$ action plane. The $\theta_\text{fast}$-oscillation of the radial phase has only a small amplitude in action space and is in perturbation studies often averaged over.

\emph{The parent orbit} closes in the frame co-rotating with the bar's pattern speed (CRF) after $l$ rotations around the bar, moving $m$-times inwards and outwards. The $m$ pericenters occur at the azimuthal positions
\begin{equation}
    \phi_{\text{CRF,peri},i} \equiv \phi_{\text{CRF,peri},0} + i\cdot \frac{2\pi l}{m}, i \in \mathds{Z}. \label{eq:all_pericenters}
\end{equation}
In angle space, the parent orbit follows therefore exactly the relation
\begin{equation}
    \theta_R(t) = -\frac{m}{l} \cdot \left(\theta_\phi(t) - \phi_\text{bar}(t) - \phi_{\text{CRF,peri},0}\right). \label{eq:thetaR_vs_thetaT}
\end{equation}
The parent orbit has a constant $\theta_\text{slow}$ and lives close to the $l$:$m$ ARL in action space. If we define the $0$-th pericenter $\phi_{\text{CRF,peri},0} \in [0,2\pi l/m]$, then it is related to the slow angle by
\begin{equation}
    \theta_\text{slow} = m \phi_{\text{CRF,peri},0}. \label{eq:slow_vs_peri}
\end{equation}
The slow angle characterizes therefore the orientation of the parent orbit with respect to the bar.

\emph{The librating orbits} are characterized by exhibiting slow oscillations around their parent orbit's $\theta_\text{slow}$. This manifests as an oscillation (i.e. \emph{libration}) around the parent orbit (i) of the pericenter and apocenter azimuthal locations with respect to the bar, (ii) in action space along a line of constant $J_\text{fast}$ with slope
\begin{equation}
    \frac{\Delta J_R}{\Delta L_z} = \frac{l}{m},
\end{equation}
and (iii) within a stripe of slope
\begin{equation}
    \frac{\Delta \theta_R}{\Delta(\theta_\phi-\phi_\text{bar})} = - \frac{m}{l}
\end{equation}
around the relation Equation \eqref{eq:thetaR_vs_thetaT} in angle space. The librating orbits differ only in their libration amplitude. The maximum possible libration amplitude depends on the strength of the bar's $m$-th Fourier mode. The boundary at the maximum libration amplitude is called the \emph{separatrix}. In action space, it is roughly parallel to the ARL on both sides and marks the region within which orbits are \emph{trapped} at the resonance and librate. Outside, orbits circulate freely.

\subsection{Examples: The angle signature of resonant orbits} \label{sec:example_trapped_orbits}

As an example, we show in the upper panels of Figure \ref{fig:example_trapped_orbits} librating orbits trapped at four different bar resonances: the 1:2 OLR, the 1:1 and 1:4 outer Lindblad resonances (1:1 and 1:4 resonances for short), and CR with $l=0$. In the frame co-rotating with the bar, they exhibit the well-known and distinctive shapes that have been discussed in depth in the literature (e.g., \citealt{1983A&A...127..349A,1989A&ARv...1..261C,2000AJ....119..800D,2001AA...373..511F,1993RPPh...56..173S,2009MNRAS.394.1605H,2018MNRAS.474.2706B,2019MNRAS.488.3324F,2020MNRAS.495..886B}). We integrated these orbits in the Galaxy potential dominated by an idealized quadrupole bar which we will introduce in Section \ref{sec:simulation_description}. For comparison, we show also one non-resonant circulating orbit.

The lower panels in Figure \ref{fig:example_trapped_orbits} present these orbits in angle space $(\theta_\phi,\theta_R)$. If we consider each orbit as the time evolution of a single star, then the $\theta_\phi$-coordinate shown on the $x$-axis is actually the co-rotating \begin{equation}
\theta_{\phi,\text{CRF}}(t) \equiv \theta_\phi(t) - \phi_\text{bar}(t) + 25^\circ.
\end{equation}
Figure \ref{fig:example_trapped_orbits} illustrates that our bar potential aligns the peri- and apocenters of resonant orbits along its major and minor axes. The slow angles of our example orbits with $l>0$ oscillate---following Equations \eqref{eq:slow_angle_def} and \eqref{eq:slow_vs_peri}--- around $\theta_\text{slow,OLR}=\pi$, $\theta_\text{slow,1:1}=3\pi/2$, and $\theta_\text{slow,1:4} = 0$. For more general bar potentials, it depends on the exact form of the potential around which values of $\theta_\text{slow}$ orbits will be trapped.

With the \emph{Gaia} data in mind, we now consider each example orbit in Figure \ref{fig:example_trapped_orbits} as a sample of many stars distributed along the same orbit and being observed at a single snapshot in time, $t_\text{today}$. $\theta_\phi$ on the $x$-axis denotes then current azimuthal angle in the coordinate frame of Figure \ref{fig:epicycle_approx_theta_phi}, where the Sun is at $\phi_0=0$ and the bar at $\phi_\text{bar} = +25^\circ$.

As a direct consequence of the libration around the parent orbit (Equation \ref{eq:slow_angle_def}), the stars on these orbits lie within stripes of slope
\begin{equation}
    \frac{\Delta \theta_R}{\Delta \theta_\phi} = -\frac{m}{l}.
\end{equation}
We expect that at the bar resonances many stars on a range of orbits accumulate along specific values of $\theta_\text{slow}$ and therefore these characteristic slopes.
The reason is that (i) the bar creates a potential trough aligned with the $m$-th bar component's axes within which the resonant stars librate such that discrete values of $\theta_\text{slow}$ are preferred (e.g. \citealt{1982ApJ...252..308B}) and (ii) the trapping regions at bar resonances (especially those with $m=2$) cover substantial regions of stellar phase-space \citep{2018MNRAS.474.2706B} trapping many stars. 
At the heart of this study---and also the works of \citet{2010MNRAS.409..145S} and \citet{2011MNRAS.418.1565M}---is therefore the idea that by searching for stellar overdensities in angle space along slopes of $-m/l$, we hope to identify the bar's $l$:$m$ resonances.

\section{An angle-based method to identify the OLR} \label{sec:method}

\subsection{The test particle simulation} \label{sec:simulation_description}

To illustrate and test this idea to identify bar resonances in \emph{Gaia}'s angle space, we have run two test particle simulations similar to those in \citet[see their appendix C for details]{2021MNRAS.500.2645T}. An axisymmetric stellar disk with 10 Million particles is generated from the disk distribution function by \citet{2011MNRAS.413.1889B} in the axisymmetric \texttt{MWPotential2014} by \citet{2015ApJS..216...29B} ($R_0 \equiv 8~\text{kpc}, v_0 \equiv  v_\text{circ}(R_0) = 220~\text{km/s}, L_{z,0} \equiv R_0 \cdot v_0 = 1760~\text{kpc km/s} , \Omega_0 \equiv v_0/R_0 = 27.5~\text{km/s/kpc}$). The pure quadrupole 3D bar model by \citet{2000AJ....119..800D} and \citet{2016MNRAS.461.3835M} is superimposed onto the background MW potential. In addition, we implemented for a $m=4$ bar component (following eq. (4) in \citealt{2018MNRAS.477.3945H}) an analogous 3D version (following eq. (1) in \citealt{2016MNRAS.461.3835M}). As definition for the strength of the different bar components, we use $\alpha_{m}$, which describes the radial force ratio at $R=R_0$ and $\phi=\phi_\text{bar}$ of the $m$-th Fourier term of the bar model vs. the axisymmetric background model. Analogously to the 2D bar model by \citet{2018MNRAS.477.3945H}, we build a 2-component bar model with $\alpha_{m=2} = 0.01$ and a weak $\alpha_{m=4} = -0.0005$. The latter makes the bar's face-on shape boxy and therefore more realistic.\footnote{Bars with pronounced non-zero $m=4$ Fourier components have been investigated in simulations (e.g., by \citealt{1981A&A....96..164C,1991A&A...252...75P,2002MNRAS.330...35A,2018MNRAS.477.3945H}), and in the dynamical MW bar model by \citet{2017MNRAS.465.1621P}. Observationally, there are indications that boxy bars might occur more often in late-type disk galaxies, and pointy bars in early-type galaxies (see, e.g., \citealt{1994ApJ...437..162Q,1996ASPC...91...37O,2006AJ....132.1859B}).} The bar model has a length of $R_\text{bar} = 4.5~\text{kpc}$ or $3.5~\text{kpc}$, and rotates with a pattern speed of $\Omega_\text{bar} = 40~\text{km/s/kpc}$ or $51~\text{km/s/kpc}$, respectively. We refer to these as the \texttt{Fiducial\_40} and the \texttt{Fiducial\_51} simulations. We slowly grow the bar strength from 0 to its final value over 5 bar periods $T_\text{bar}\equiv 2\pi/\Omega_\text{bar}$. All particle orbits are integrated in this barred galaxy potential for $T_\text{int} = 50T_\text{bar}$ using \texttt{galpy}\footnote{The Python package \texttt{galpy} for Galactic Dynamics by \citet{2015ApJS..216...29B} lives at \url{http://github.com/jobovy/galpy}.}. The azimuth of the bar's major axis in the simulation is 
\begin{equation}
    \phi_\text{bar} = \phi_{\text{bar},0} +  \Omega_\text{bar}t, \hspace{1cm}\phi_{\text{bar},0} \equiv 25^\circ.
\end{equation}
After each full period of orbit integration it is therefore similar to today's configuration between the MW's bar and the Sun, where $\phi_0 = 0$. Snapshots at $t=25T_\text{bar}$ and $t=50T_\text{bar}$ are stacked to increase the number of particles to 20 Million. To this final sample, we apply cuts $L_z \in [0.4,1.5]L_{z,0}$ and $J_z <0.01 L_{z,0}$.

\begin{figure}
    \centering
    \includegraphics[width=\columnwidth,trim={10 0 110 20}, clip]{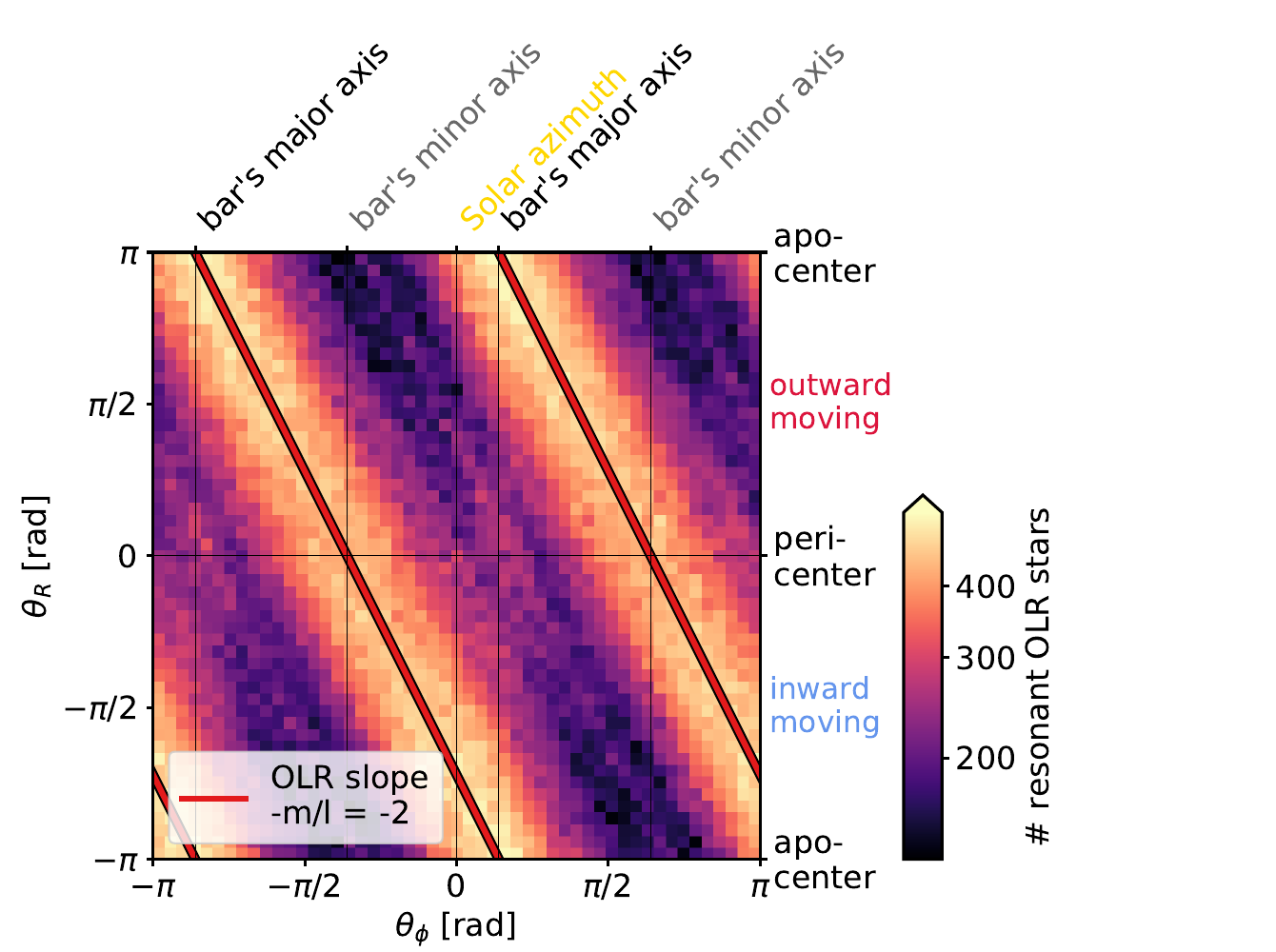}
    \caption{Angle space $(\theta_\phi,\theta_R)$ distribution of stars in the \texttt{Fiducial\_40} simulation that satisfy the resonance condition in Equation \eqref{eq:select_resonant_stars} for $l=+1, m=2$, i.e. are at the bar's Outer Lindblad Resonance (OLR). The characteristic slope of $\Delta \theta_R / \Delta \theta_\phi = -2$ might help to identify the real Galactic bar's OLR in the \emph{Gaia} data.}
    \label{fig:OLR_resonant_stars}
\end{figure}

The simulations contain only stars on loop disk orbits that are rotating in the inertial frame in $\phi$ and oscillating in $R$. In the \texttt{Fiducial\_40} simulation, we can therefore determine  the dominant orbital frequencies $\Omega_{\phi,\text{true}}$ and $\Omega_{R,\text{true}}$ from the maximum peak of the Fourier transform of the star's time evolution in $R(t)$ and $\cos\phi(t)$ (e.g. \citealt{1982ApJ...252..308B,1993CeMDA..56..191L,2019MNRAS.488.3324F} and Figure \ref{fig:slow_angle_frequencies}) with a precision of $2\pi/(T_\text{int})=2\pi/(50T_\text{bar})=0.8~\text{km/s/kpc}$. We select stars in the simulation that are (not in the axisymmetric estimate but) actually in the barred system in resonance with the bar's OLR ($l=1,m=2$) using the resonance condition
\begin{equation}
\begin{aligned}
    \mid \Omega_\text{bar} - \Omega_{\phi,\text{true}} - \frac{l}{m} \times \Omega_{R,\text{true}} \mid &< 0.8~\text{km/s/kpc}.
    \end{aligned}
    \label{eq:select_resonant_stars}
\end{equation}

Their distribution peaks around the parent orbits (Equation \ref{eq:thetaR_vs_thetaT}) that all follow the ($2\pi$-wrapped) line
\begin{eqnarray}
    \theta_{R,x_1(1)} &=& -2 (\theta_\phi - \phi_\text{bar}) + \theta_\text{slow,$x_1(1)$},\\
\text{with } \theta_\text{slow,$x_1(1)$} &\equiv& \pi
\end{eqnarray}
(c.f. \citealt[eq. 39]{2018MNRAS.474.2706B}), with a slope of $-m/l=-2$ and with pericenter at the bar's minor axis. These orbits belong to the class of $x_1(1)$ OLR orbits aligned with the bar. Some OLR stars are found in Figure \ref{fig:OLR_resonant_stars} on orbits that have their pericenters close to the bar's \emph{major} axis; in our simulation, these exhibit low libration and belong to the anti-aligned $x_1(2)$ orbit class, which lives inside of the OLR ARL. Its parent orbits follow lines 
\begin{eqnarray}
    \theta_{R,x_1(2)} &=& -2 (\theta_\phi - \phi_\text{bar}) + \theta_\text{slow,$x_1(2)$},\\
\text{with } \theta_\text{slow,$x_1(2)$} &\equiv& 0.
\end{eqnarray}

Figure \ref{fig:OLR_resonant_stars} confirms for the OLR that---as suggested in Section \ref{sec:example_trapped_orbits}---a stellar sub-selection, which contains stars on a range of true $l$:$m$ resonant orbits, creates angle overdensities along stripes with the characteristic slope of $-m/l$.

\subsection{Selecting resonant stars} \label{sec:selecting_resonant_stars}

No clear signal of resonances is visible in angle space when considering \emph{all} stars at once in either the simulation or the \emph{Gaia} data, if no pre-selection on orbits is performed. Two aspects contribute to this:  (i) Stars at different bar resonances live at different $L_z$ and radius, but can populate overlapping regions in angle space (Figure \ref{fig:example_trapped_orbits}) (ii) Non-resonant circulating orbits smoothly `pollute' the whole $(\theta_\phi,\theta_R)$ plane, decreasing the contrast of the resonant signal.

The studies by \citet{2019MNRAS.484.3154S} and \citet{2019MNRAS.490.1026H} show that \emph{Gaia} DR2's angle space for the Solar neighbourhood $d < 200~\text{pc}$ exhibits strong sub-structure.\footnote{The reason that the angle plane for $d < 200~\text{pc}$ has a lot of substructure is the following: The in-plane phase space $(R\sim R_0,\phi\sim \phi_0,v_R,v_\phi)$ is here essentially 2D. Studying $(v_R,v_\phi)$ isolates the orbit structure therefore very cleanly. For this small volume, any 2D projection of the same phase-space, e.g. to $(L_z,J_R)$ or $(\theta_\phi,\theta_R)$, reveals the same overdensities. This can be well seen in figs. 2-3 by \citet{2019MNRAS.490.1026H}.} The angle plane for all stars out to $d = 3~\text{kpc}$ is featureless because of the above mentioned reasons. A sub-selection on orbits is therefore crucial to isolate and reveal resonant signatures in angle space. The action plane $(L_z,J_R)$ will help with this, because---as is very well-known---resonant stars create global ridges in action space (see, e.g., \citealt{2002MNRAS.336..785S,2012ApJ...751...44S,2018MNRAS.474.2706B}).

\begin{figure}
    \centering
    \includegraphics[width=\columnwidth]{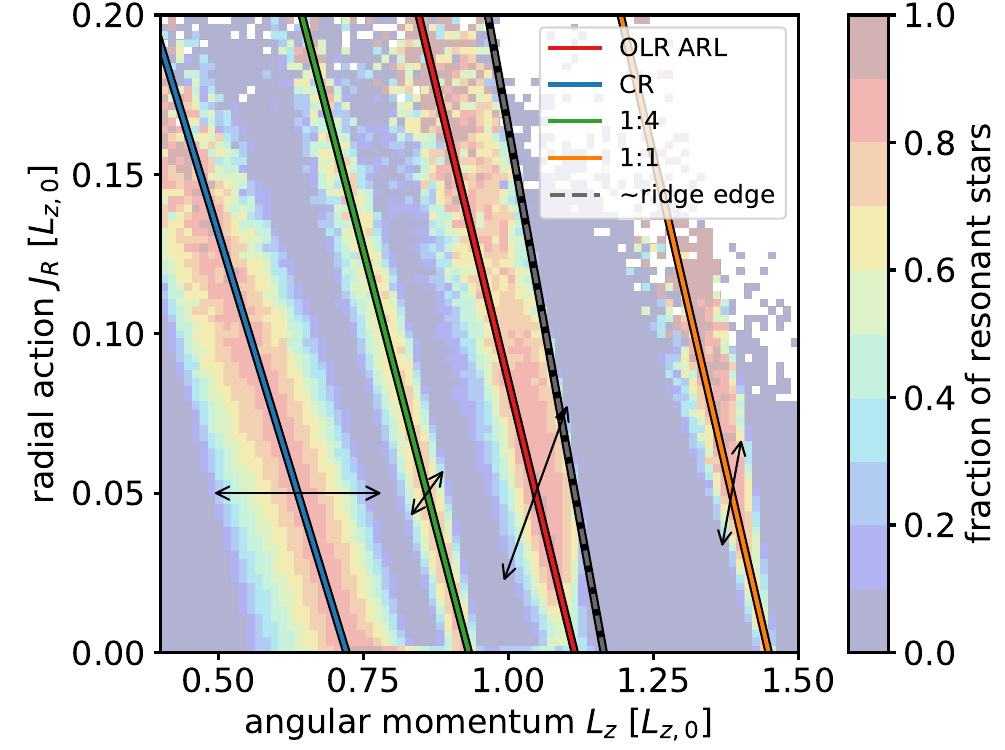}
    \caption{Fraction of resonant stars to the total number of stars in the \texttt{Fiducial\_40} simulation as a function of action space $(L_z,J_R)$. Resonant stars are selected using the resonance condition in Equation \eqref{eq:select_resonant_stars} for $l$:$m =$ 0:1, 1:4, 1:2, 1:1. Overplotted are the corresponding ARLs. The dashed line with slope $\Delta J_R/\Delta L_z = -1$ follows roughly the edge of the OLR's scattering ridge. OLR resonant stars can be found in the ridge; 1:4 resonant stars are mixed with the background for the weak $m=4$ bar component used in this work. The black arrows illustrate the direction and maximum amplitude of libration around the ARLs.}
    \label{fig:fraction_of_resonant_stars}
\end{figure}

Using Equation \eqref{eq:select_resonant_stars}, we select in the simulation all particles whose orbital frequencies satisfy the resonance condition for either CR, OLR, outer 1:1, or 1:4 resonance. Figure \ref{fig:fraction_of_resonant_stars} shows the fraction of these resonant stars with respect to all stars in the simulation. Overplotted are also the corresponding ARLs.

We confirm that the high-$J_R$ ridge to the right of the 1:2 OLR and the 1:1 ARLs (c.f. \citetalias{2021MNRAS.500.2645T}) indeed contain a high fraction of resonant stars. We call these action overdensities made up from resonant stars `scattering ridges' (see also \citetalias{2021MNRAS.500.2645T}) to describe that \emph{on average} the stellar population has been displaced towards higher $J_R$ at and by this resonance, as compared to the smooth initial disk. The width of the ridge depends on the libration strength around the ARL.

We will now focus on the 1:2 OLR, which gives the strongest signal, but discuss the other resonances briefly in Section \ref{sec:curious_coincidence} and Appendices \ref{app:11_resonance}-\ref{app:14_resonance}.

 The high-$L_z$ edge of the OLR ridge corresponds to the outer boundary of the zone of entrapment \citep{2018MNRAS.474.2706B}. Figure \ref{fig:fraction_of_resonant_stars} illustrates this by overplotting the maximum libration amplitude at $J_R\sim0.5~L_{z,0}$. We mark this edge of the OLR ridge with a line (the `ridge edge line', REL), for which we chose the slope $\Delta J_R/\Delta L_z = -1$ by eye.\footnote{At low $J_R$, this choice of REL is not a good description for the real zone of entrapment and the separatrix at the OLR, as can be seen in Figure \ref{fig:fraction_of_resonant_stars} and as discussed in \citet{2020MNRAS.495..886B}, but we found that for our practical purposes approximating it by a straight line is sufficient.} The majority of action ridges observed in the \emph{Gaia} data by \citetalias{2019MNRAS.484.3291T} have slopes steeper than the OLR ARL, but shallower than this REL. In addition, we shift the REL by $\Delta L_z$ with respect to the OLR ARL; in the simulation we chose $\Delta L_z = +0.05 L_{z,0}$ at $J_R=0$. Using the REL and the ARL as boundaries for the selection will provide a stellar sub-sample with a high fraction of true OLR stars.

\begin{figure*}
    \centering
    \includegraphics[width=\textwidth]{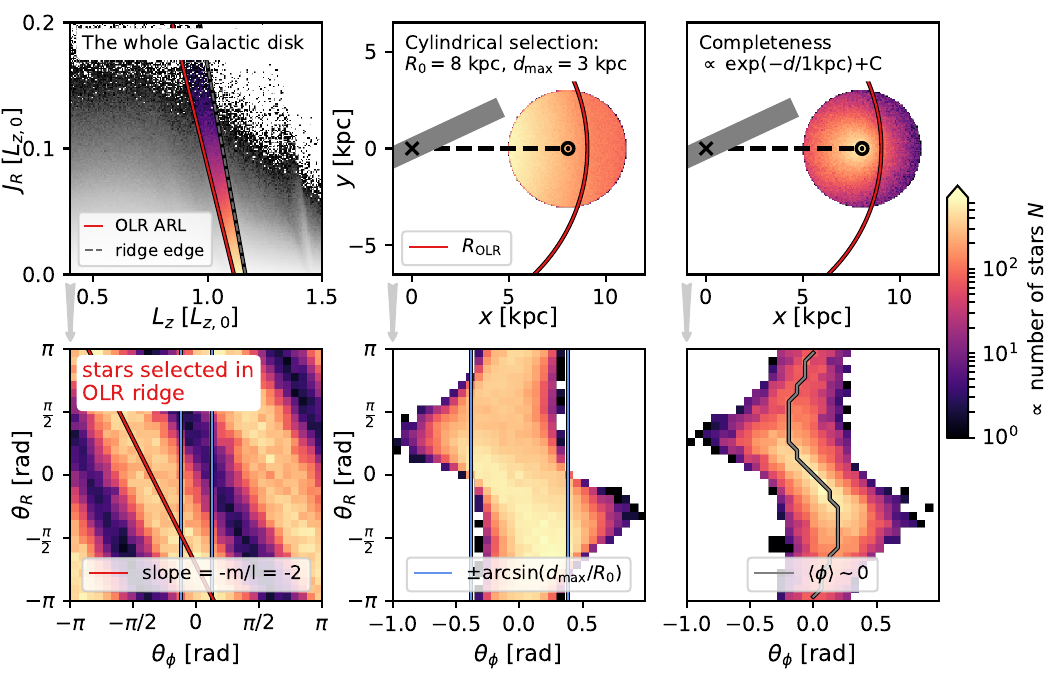}
    \caption{Illustration of survey selection effects in angle space (in the \texttt{Fiducial\_40} simulation). \emph{1st column:} The characteristic OLR slope of $-2$ in angle space is still clear when selecting stars directly from the scattering ridge in action space for the whole disk. \emph{2nd column:} Restricting the data to a cylinder around the Sun reduces the $\theta_\phi$ range substantially. \emph{3rd column:} An additional completeness function is imposed that decreases with distance. The stellar density peaks now at the Solar azimuth $\phi = 0$, corresponding to an S-shaped curve in angle space (see text for details). This selection effect hides the OLR signature.}
    \label{fig:angles_selection_effects_explained}
\end{figure*}

When creating Figure \ref{fig:fraction_of_resonant_stars}, we knew the true barred potential and therefore the true orbital frequencies. Neither is available for the MW, so a selection of resonant star candidates can in reality not be improved beyond selecting \emph{all} stars between the OLR ARL and the REL for an assumed $\Omega_\text{bar}$. In the left column of Figure \ref{fig:angles_selection_effects_explained}, we demonstrate that this proposed selection in action space (upper left panel) works well to isolate enough stars on OLR $x_1(1)$ orbits to make the OLR's angle signature of slope -2 and pericenter at $\theta_{\phi,\text{peri}}=\theta_\text{slow}/m + \phi_\text{bar}$ visible (lower left panel).\footnote{If we knew the true $\Omega_\text{bar}$ of the MW, varying the REL slope and $\Delta L_z$ could be used as a constraint on the \emph{strength} of the bar.}

\begin{figure*}
    \centering
    \includegraphics[width=\textwidth]{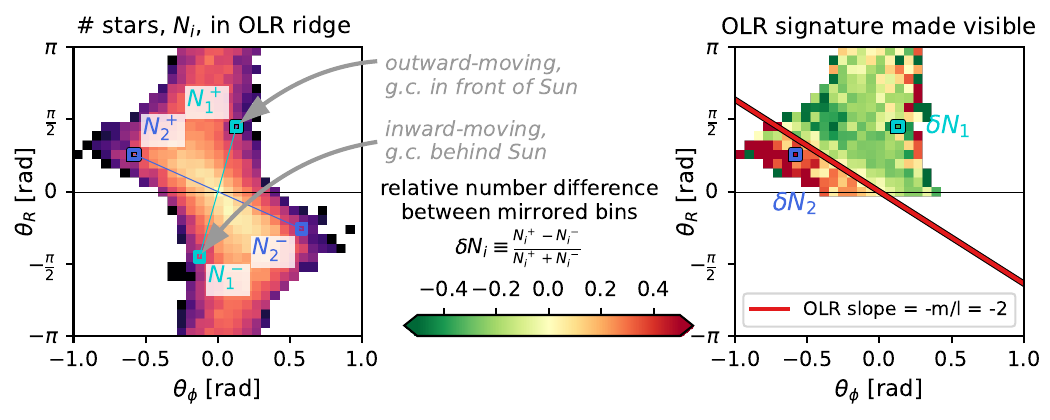}
    \caption{Dealing with selection effects: Completeness decreasing with distance from the Sun (in the \texttt{Fiducial\_40} simulation). \emph{Left:} Stars from the OLR ridge in angle space (same as lower right panel in Figure \ref{fig:angles_selection_effects_explained}). We compare the number of stars at $N_i^+ \equiv (\theta_\phi,\theta_R)$ with those at $N_i^- \equiv (-\theta_\phi,-\theta_R)$. \emph{Right:} colour-coding angle space ($\theta_R > 0$) by the relative stellar number difference makes the OLR slope visible.}
    \label{fig:angle_mirroring_explained}
\end{figure*}

\begin{figure*}
    \centering
    \includegraphics[width=\textwidth]{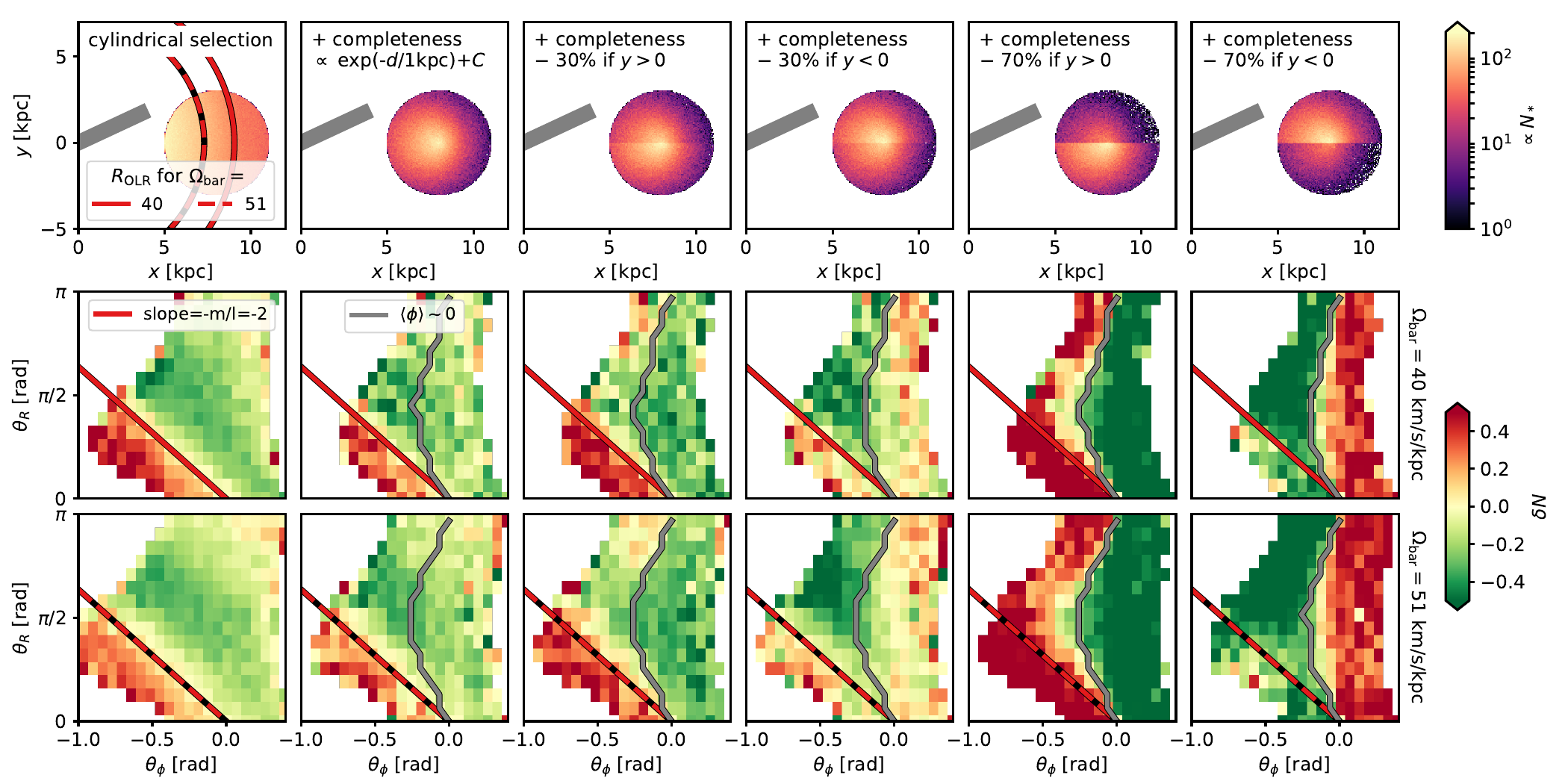}
    \caption{Dealing with selection effects: Completeness variation with Galactic longitude. For two simulations with different bar pattern speeds (\texttt{Fiducial\_40} and \texttt{Fiducial\_51}), it is shown how the clean OLR signature in relative star numbers $\delta N$ (\emph{1st and 2nd column}) is still visible for completeness differences of up to 30\% (\emph{3rd and 4th column}), but is overtaken by selection effects for completeness differences of 70\% (\emph{5th and 6th column}). In the latter case, the pattern in $\delta N$ follows the shape of the selection-function-affected stellar number distribution (denoted by the grey line along which $\langle \phi \rangle \sim 0$) rather than the OLR slope (red line).}
    \label{fig:issues_with_incompleteness}
\end{figure*}

\subsection{Dealing with selection effects around the Sun} \label{sec:selection_effects_with_distance}

If we had access to stars in the whole Galactic disk, $\phi \in [-\pi,\pi]$ (as in the first column of Figure \ref{fig:angles_selection_effects_explained}), we could not only identify the bar's real OLR and pattern speed from the sloped stripes in angle space, but their location in $\theta_\phi$ would also reveal the bar's orientation with respect to the Sun.

Unfortunately, \emph{Gaia} provides currently only precise data around the Sun, $\vect{x}_\odot = (R_0,\phi_0,z_0) = (8~\text{kpc},0^\circ,25~\text{pc})$, out to distances of roughly $d_\text{max}\sim3~\text{kpc}$, even when using improved distance estimates by, e.g., \citet{2019MNRAS.487.3568S}. As demonstrated in the second column of Figure \ref{fig:angles_selection_effects_explained}, this reduces the observable $\theta_\phi$ range within the Galaxy to \begin{equation}
|\theta_\phi| \sim |\phi|< \arcsin \left( \frac{d_\text{max}}{R_0} \right).
\end{equation}
Near-circular orbits crossing this survey volume can be observed at any radial phase $\theta_R$. For orbits with high $J_R$, only parts of their epicycles enter the survey volume. If a star is observed close to the Sun, but its guiding center is azimuthally in front of the Sun ($\theta_\phi > 0$), it will currently be inward-moving ($\theta_R < 0$; see Figure \ref{fig:angles_in_epicycle_approximation_explained}); if a star is observed while its guiding center is behind the Sun ($\theta_\phi < 0$), it will be outward-moving ($\theta_R > 0$). These high-$J_R$ stars make up the $|\theta_\phi| >  \arcsin \left(d_\text{max}/R_0 \right)$ wings in the corresponding angle distribution. How extended these wings are depends on the population properties, in particular on the radial velocity dispersion $\sigma_R$. If the data are complete within this survey volume, density variations along the OLR slope are still discernible in angle space.

However, the completeness of surveys depends also on the apparent brightness of the stars (e.g. \citealt{2020MNRAS.497.4246B} for \emph{Gaia} DR2). For simplicity, we assume the completeness decreases with distance from the Sun $d$ as
\begin{equation}
    \text{completeness}(d) \propto \exp \left(-\frac{d}{1~\text{kpc}}\right) + \text{const.} \label{eq:radial_incompleteness}
\end{equation}
We apply this to the data from the \texttt{Fiducial\_40} simulation in Figure \ref{fig:angles_selection_effects_explained}, last column. 

To understand how the incompleteness shapes the angle distribution, it is convenient to mark the location in the $(\theta_\phi,\theta_R)$ plane at which the average azimuth of the stars $\langle \phi \rangle \sim 0$. This line runs (in an almost axisymmetric disk, and as a function of $\theta_R$) along the peak of the stellar number distribution, as can be seen in the lower right panel of Figure \ref{fig:angles_selection_effects_explained}. The `$\langle \phi \rangle \sim 0$' line and the peak line agree for this completeness function by construction, because most stars are observed at the Solar azimuth $\phi = 0$ (at a given $R$). The S-shape of the peak `$\langle \phi \rangle\sim0$' line in Figure \ref{fig:angles_selection_effects_explained} can be explained as follows: At peri- and apocenter, $\theta_R \in \{ \pm \pi,0\}$, $\theta_\phi$ corresponds to the real $\phi$ of the star (c.f. \citealt[fig. 1]{2011MNRAS.418.1565M}). Consequently, at these $\theta_R$s, the maximum number of stars is observed at $\phi=\theta_\phi=0$. The many stars that are currently at $\phi=0$, but have $\theta_R$ around $+\pi/2$ or $-\pi/2$, have guiding centers further away from the Sun, at $\theta_\phi < 0$ or $\theta_\phi > 0$, respectively. Here $\theta_\phi \neq \phi$, and the exact location of the density peak depends on the radial velocity dispersion. By going away from this peak line in $\theta_\phi$, the number density decreases (and $|\langle \phi \rangle|$ increases) smoothly towards the wings of the distribution.

A pedagogic explanation of the selection effects in angle space can also be found in the appendix of \citet{2011MNRAS.418.2459H}.

For this completeness function, the physical angle signature of the OLR is hidden by selection effects in the lower right panel of Figure \ref{fig:angles_selection_effects_explained}. If we knew the real form of \emph{Gaia}'s incompleteness function (like Equation \eqref{eq:radial_incompleteness} for the \texttt{Fiducial\_40} simulation), it would be possible to perform full-likelihood forward-modeling of the angle distribution to uncover the OLR signature (c.f. also \citealt{2011MNRAS.418.1565M}).

Here, we present a simpler strategy to make the OLR's angle signature visible that does not require precise knowledge of the selection function. We make use of the fact that an incompleteness function like that presented in Equation \eqref{eq:radial_incompleteness} affects outward- and inward-moving stars in equal measures, leading to the symmetric S-shaped envelope in angle space. We mirror the angle distribution twice---horizontally at $\theta_\phi=0$ and vertically\footnote{The results do not change if the mirroring in $\theta_R$ is performed at pericenter or at apocenter.} at $\theta_R=0$---and calculate the \emph{relative} number difference between the original and the mirrored angle distribution. By doing so, we effectively \emph{remove} the incompleteness decrease at $|\phi| > 0$. Physically, this corresponds to comparing the number of outward moving stars with the same guiding center location $\theta_\phi$,
\begin{equation}
    N_i^+ \equiv N_i(\theta_{\phi,i},\theta_{R,i})
\end{equation}
with the number of inward-moving stars that have guiding center locations on the opposite $\phi$-side with respect to the Sun, i.e.
\begin{equation}
N_i^- \equiv N_i(-\theta_{\phi,i},-\theta_{R,i}).
\end{equation}
As both bins should be affected in the same way by selection effects, we scale the difference by the total number of stars in these two bins. We illustrate this strategy in Figure \ref{fig:angle_mirroring_explained}. The relative residuals,
\begin{equation}
    \delta N_i \equiv \frac{N_i^+ - N_i^-}{N_i^+ + N_i^-}, \label{eq:delta_N}
\end{equation}
reveal now the OLR signature visually and clearly by transitioning through zero along the line
\begin{equation}
    \theta_R = -2 \theta_\phi,
\end{equation}
which we call OAS ("OLR angle slope" line). The OAS line goes through $(\theta_\phi,\theta_R)=(0,0)$ by construction, and the region $\theta_R < 0$ does not contain additional information because of the mirroring. For $\theta_R>0$, the distribution below the OAS line has $\delta N > 0$ (`red' = dominated by outward-moving stars), and above $\delta N < 0$ (`green' = dominated by inward-moving stars). This tells us that we are looking at $x_1(1)$ orbits. For $x_1(2)$ orbits, it would be the other way round.

The width of the `green' stripe above the OAS line in principle still contains some information about the bar orientation angle and the width of the unmirrored stripe due to the libration strength of the OLR orbits, but requires careful modeling. In this work, we will just focus on investigating the pattern in $\delta N$ around the OAS line.

\subsection{Dealing with selection effects due to asymmetric sky coverage} \label{sec:selection_effects_with_longitude}

In the previous section, we showed how the OLR's angle signature can be made visible in the case of a symmetric incompleteness function with respect to the Sun. In particular, the symmetry in Galactic longitude---$\pm l$ with respect to the Galactic center---is crucial for our method to work: Stars with the same $|\theta_R|, |\theta_\phi|$, and $(L_z,J_R)$ live at the same $\pm \phi$ and thus $\pm l$. The completeness symmetry in $l$ is therefore more important for obtaining a physically meaningful $\delta N$ (Equation \ref{eq:delta_N}) than the absolute value of the completeness.

The assumption of longitude symmetry might, however, be too simplified for real MW data. \citet{2020MNRAS.497.4246B} investigated \emph{Gaia} DR2's completeness by modeling the survey's sky coverage using \emph{Gaia}'s scanning law. The latter is not perfectly symmetric with respect to the Galactic center. \citet{2021MNRAS.500..397R} constructed an empirical completeness function of the \emph{Gaia} DR2 RVS sub-sample with respect to the full DR2 catalogue. Within the magnitude range 
\begin{equation}
    5 <G_\text{RVS}/\text{mag} <12 \label{eq:mag_cut}
\end{equation}
and colour range 
\begin{equation}
0.35<(G-G_\text{RP})/\text{mag}<1.25 \label{eq:colour_cut}
\end{equation} this internal completeness is $80\%$ or better (see fig. 3 in \citealt{2021MNRAS.500..397R}). The internal completeness varies with position on the sky. By comparing the completeness function\footnote{A Python tutorial for the \emph{Gaia} DR2 RVS selection function by \citet{{2021MNRAS.500..397R}} can be found at \myurlrvs.} by \citet{2021MNRAS.500..397R} at locations $+l$ vs. $-l$, the differences are mostly lower than 15\% across the sky. An exception are the regions $|l|\in [ 10^\circ,30^\circ]$, $b\in[-20^\circ,+20^\circ]$, where the completeness differences can be up to $\sim 30\%$.

\citet{2021MNRAS.500..397R} argue that the \emph{internal} completeness of \emph{Gaia} RVS can be considered as an \emph{external} completeness as well. \citet{2018ascl.soft11018R} have investigated\footnote{The tutorial by \citet{2018ascl.soft11018R} showing the completeness of \emph{Gaia} DR2 with respect to 2MASS, can be found at \myurl.} that \emph{Gaia} DR2 itself is almost complete with respect to 2MASS \citep{2006AJ....131.1163S} in the magnitude range $ 8 < G/\text{mag} < 12$. Completeness differences of up to $\sim 30\%$ can therefore be considered as a realistic value also for the \emph{Gaia} RVS sample with respect to the true stellar distribution.

To investigate if such completeness differences in $\pm l$ can hide the physical OLR signature in angle space, we remove 0\%, 30\%, or 70\% of the test particles at Galactic coordinates $y>0$ or $y<0$ from both the \texttt{Fiducial\_40} and \texttt{Fiducial\_51} simulations. Figure \ref{fig:issues_with_incompleteness} compares the corresponding $\delta N$ distributions with the OAS line and the `$\langle \phi \rangle \sim 0$' line in angle space.

We find that for 70\% completeness differences the `red/green' pattern follows the `$\langle \phi \rangle \sim 0$' line because of the imposed completeness break at $\phi=0$. We conclude from this that if the $\delta N$ distribution follows closely the `$\langle \phi \rangle \sim 0$' line, there is a danger that what we are seeing is not a physical signature, but rather a selection effect.

In the case of 30\% completeness differences, the pattern in $\delta N$ still follows cleanly the OAS line, just as in the ideal case of 0\% completeness difference. We conclude therefore that the completeness variation across $\pm l$ in \emph{Gaia} DR2 RVS are not strong enough to hide the physical resonance signatures in angle space.

\section{Results from the \emph{Gaia} data}

\subsection{\emph{Gaia} DR2 RVS data} \label{sec:data_description}

We use the \emph{Gaia} DR2 \citep{2018A&A...616A...1G} sub-sample that has, in addition to on-sky positions and proper motions, also radial velocity measurements from \emph{Gaia}'s Radial Velocity Spectrometer (RVS, \citealt{2019A&A...622A.205K}). We clean this sample by applying the cuts in brightness and colour (Equations \ref{eq:mag_cut}-\ref{eq:colour_cut}) proposed by \citet{2021MNRAS.500..397R}. For the distances to the stars, we use the Bayesian estimates by \citet{2019MNRAS.487.3568S} and cut at $d \leq 3~\text{kpc}$ to ensure good accuracy and precision. 

We transform to Galactocentric coordinates using the Solar peculiar motion measurements by \citet{2010MNRAS.403.1829S}, $R_0 = 8~\text{kpc}$ and $v_\text{circ}(R_0) = 220~\text{km/s}$ in the gravitational \texttt{MWPotential2014} model by \citet{2015ApJS..216...29B}, as well as $z_0 = 25~\text{pc}$ \citep{2008ApJ...673..864J}. Axisymmetric actions, angles, and frequencies are estimated in the \texttt{MWPotential2014} using the St\"{a}ckel fudge algorithm \citep{2012MNRAS.426.1324B}, just as for the test particles in the simulation, and as done for the \emph{Gaia} data in \citetalias{2019MNRAS.484.3291T} and \citetalias{2021MNRAS.500.2645T}. A further cut is imposed to focus on orbits with small vertical excursions close to the Galactic plane ($J_z < 0.01L_{z,0}$).

We present \emph{Gaia}'s full 4D action-angle space in Appendix \ref{app:Gaias_full_angle_space_MWPot} in Figure \ref{fig:angle_space_scan_OLR}.

As we will show in Appendix \ref{app:Eilers_pot}, the details of the assumed gravitational MW potential do not matter for the subsequent analysis. In particular, we have confirmed the qualitative results that we will present in the next sections also for the MW model by \citet{2019ApJ...871..120E}.

Using \emph{Gaia} EDR3 \citep{2021A&A...649A...1G} rather than DR2 would also not affect our results. Proper motion uncertainties $\delta\mu$ are by a factor 2-3 smaller in EDR3 than in DR2. For the previously already very precisely measured DR2 proper motions for bright stars with $\delta\mu\sim0.06~\text{mas yr}^{-1}$---like those in the RVS sample---this improvement does not have a significant impact. \emph{Gaia} parallaxes have not been used in this work.

\subsection{Candidates for the MW bar's OLR in angle space} \label{sec:OLR_candidates}

\begin{figure*}
    \centering
    \includegraphics[width=\textwidth]{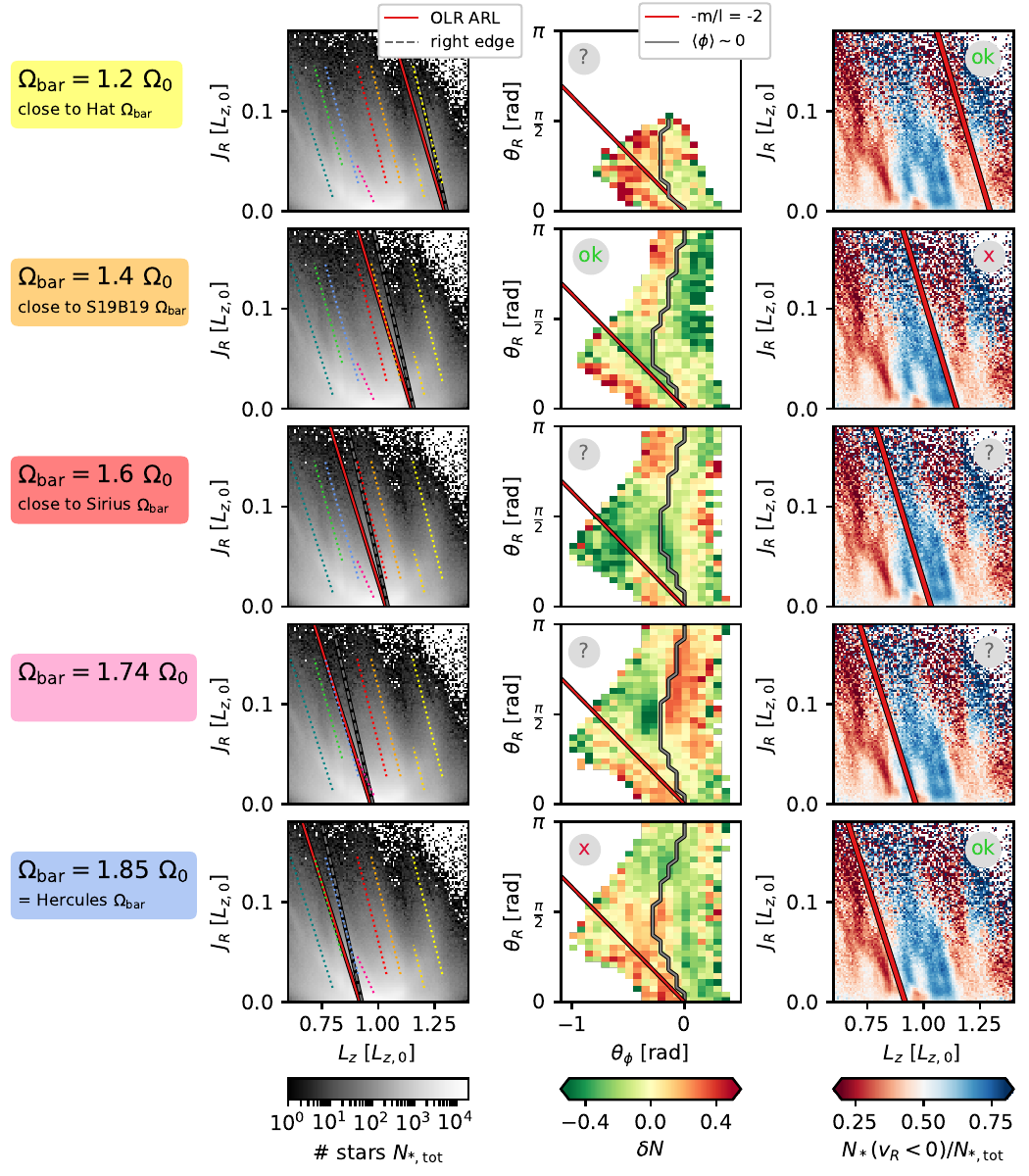}
    \caption{OLR candidates in the \emph{Gaia} DR2 action-angle data. \emph{Middle:} Four possible pattern speed candidates were identified from \emph{Gaia}'s full angle space in Figure \ref{fig:angle_space_scan_OLR} and are now presented here: $\Omega_\text{bar} \in \{1.2,1.4,1.6,1.74\} \Omega_0$. The corresponding angle planes show that the red OAS separates a `red' from a `green' stripe in $\delta N$, as expected for the OLR. Interestingly, the first three candidates agree with favourites in the literature. \emph{Left:} In the full action space, the red OLR ARL and grey dashed line mark the respective sub-selection of stars shown in angle space. The dotted lines denote the moving group ridges following \citetalias{2019MNRAS.484.3291T}, e.g., the \texttt{Hat (I, yellow)}, \texttt{Sirius (G1, orange)}, and \texttt{Hercules/Horn (D1, blue)}. \emph{Right:} In addition, we show the location of the OLR ARL on top of the $v_R$-asymmetry in $(L_z,J_R)$. A 'red/blue` signature (c.f. \citetalias{2021MNRAS.500.2645T} and references therein) would be expected around the true OLR. Because of selection effects in angle space at the \texttt{Hat}, or not fully understood mechanisms causing \texttt{Sirius}, no clear preference for either the \emph{slow} \texttt{Hat} or \emph{slightly faster slow} \texttt{S19B19} $\Omega_\text{bar}$ can be reached. The \emph{fast} \texttt{Hercules} \emph{bar} pattern speed, shown for comparison at the \emph{bottom}, can, however, be ruled out by the angle distribution.}
    \label{fig:angle_space_candidate_OLR}
\end{figure*}

Based on the resonant action-angle-signature presented in Section \ref{sec:method}, we will now search in the \emph{Gaia} data for the bar's true OLR. For this, we adopt the following independent approach, which is also illustrated in Figure \ref{fig:angle_space_scan_OLR} in Appendix \ref{app:Gaias_full_angle_space_MWPot}.

\begin{enumerate}[leftmargin=*,label=Step \arabic*:]
\item Picking a bar pattern speed $\Omega_\text{bar}$, and finding the location of the corresponding OLR ARL in action space $(L_z,J_R)$.
\item Assuming that a large fraction of true OLR stars is found to the right of the ARL, and therefore selecting stars between the ARL and an assumed REL with slope $-1$ and shifted by $\Delta L_z = +0.01L_{z,0}$ with respect to the ARL at $J_R=0$.\footnote{For the test particle simulations we have used $\Delta L_z = 0.05L_{z,0}$ as compared to $\Delta L_z = 0.01L_{z,0}$ in the \emph{Gaia} data. As can be seen in Figure \ref{fig:fraction_of_resonant_stars}, also a smaller action area to the right of the ARL would contain many OLR stars. The \emph{Gaia} data are so plentiful that also this smaller area provides excellent number statistics.}
\item Comparing the $\delta N(\theta_\phi,\theta_R)$ distribution of this stellar sub-selection with the model prediction for true OLR $x_1(1)$ orbits in the simulation (Figure \ref{fig:angle_mirroring_explained} and lower left in Figure \ref{fig:angle_space_scan_OLR}) along the OAS line of slope -2 to identify if this could be a candidate for the real OLR---and therefore for the bar pattern speed.
\item Comparing $\delta N(\theta_\phi,\theta_R)$ also to the `$\langle \phi \rangle \sim 0$' line to check if selection effects might `fake' the OLR signature (see Figure \ref{fig:issues_with_incompleteness}).
\item Repeating Steps 1-4 for varying $\Omega_\text{bar}$. By doing so, we effectively scan with the OLR ARL across the action plane (upper left panel in Figure \ref{fig:angle_space_scan_OLR}) and study the corresponding angle planes (right panels in Figure \ref{fig:angle_space_scan_OLR}).
\end{enumerate}

The structures in the \emph{Gaia} data are complex---also in angle space. A meaningful quantitative criterion to decide which angle plane is the one and only true bar OLR is therefore beyond our current ability to understand and model the data. We restrict ourselves therefore to a qualitative (and therefore disputable) by-eye identification of possible candidates. The reader is encouraged to read-off in Figure \ref{fig:angle_space_scan_OLR} the $\Omega_\text{bar}$ for which they themselves think the data look closest to the model prediction.

For the casual reader, we recommend to jump directly to Figure \ref{fig:angle_space_candidate_OLR}, where we present four independently identified OLR candidates that we think look convincing: $\Omega_\text{bar} \in \{1.2, 1.4, 1.6, 1.74\} \Omega_0$.

\begin{itemize}[leftmargin=*,topsep=1ex,itemsep=1ex]
    \item $\Omega_\text{bar} = 1.2 \Omega_0 = 33~\text{km/s/kpc}$ associates the \texttt{Hat (I, yellow)} action ridge with the OLR and is interestingly very similar to \emph{slow bar} models deduced from the local velocities in the literature (c.f., e.g., \citealt{2019AA...626A..41M,2020MNRAS.495..895B}; \citetalias{2021MNRAS.500.2645T}; \citealt{2021MNRAS.508..728K}). Unfortunately, the angle data are noisy and restricted to close-to-pericenter stars and the `$\langle \phi \rangle \sim 0$' line overlaps with the OAS line, suggesting that we might see selection effects here rather than a physical signature.
    \item $\Omega_\text{bar} = 1.4 \Omega_0 = 38.5~\text{km/s/kpc}$ suggests that the \texttt{Sirius (G1, orange)} ridge is close to the OLR (c.f. \citealt{2021MNRAS.508..728K}). This pattern speed is in our opinion the strongest candidate in angle space. Surprisingly, this measurement is consistent with direct measurements of the bar pattern speed in the Galactic center (c.f. \citealt{2017MNRAS.465.1621P,2019MNRAS.488.4552S,2019MNRAS.489.3519C,2019MNRAS.490.4740B,2017ApJ...840L...2P}).
    \item $\Omega_\text{bar} = 1.6 \Omega_0 = 44~\text{km/s/kpc}$---with the OLR located between the \texttt{Hyades/Horn (E1, pink)} and the \texttt{Sirius (F1, red)} ridges---is a weak candidate because of the large inward-moving (`green') blob at $\theta_R\sim\pi/4$. But we point out that at least at low $\theta_R$ the $\delta N$ pattern follows exactly the OAS, while the `$\langle \phi \rangle \sim 0$' line has a very different slope, suggesting that this signature is physical. This pattern speed is close to suggestions by \citet{1991dodg.conf..323K}, \citet{2019MNRAS.490.1026H}, \citetalias{2021MNRAS.500.2645T}, and \citet{2021MNRAS.508..728K} from the local action space, and to measurements from the Galactic center, $41\pm3~\text{km/s/kpc}$ \citep{2019MNRAS.488.4552S,2019MNRAS.490.4740B}.
    \item $\Omega_\text{bar} = 1.74 \Omega_0 = 47.9~\text{km/s/kpc}$ aligns the OLR ARL with the `Pleiades', the `Hyades' and the `Horn' (the \texttt{Hyades/Horn (D1, pink)} ridge in action space) and the signature along the OAS in angle space is distinct from possible selection effects.
\end{itemize}

Even the last pattern speed is slower than the classical and well-constrained \emph{fast bar} pattern speed of $\Omega_\text{bar}=1.85\Omega_0=51~\text{km/s/kpc}$ in the literature \citep{2000AJ....119..800D,2014AA...563A..60A} that associates the `Hercules/Horn' bimodality with the OLR (see also, e.g., \citealt{2019MNRAS.488.3324F}; this corresponds to the \texttt{Hercules (C1, green)} and \texttt{Hercules/Horn (D1, blue)} ridges in action space). For comparison, we show in the bottom row of Figure \ref{fig:angle_space_candidate_OLR} also the angle space for this \texttt{Hercules} pattern speed. In this region of phase-space, the `$\langle \phi \rangle\sim0$' line does not follow the observed `red/green' transition in $\delta N$; based on Section \ref{sec:selection_effects_with_longitude}, selection effects are therefore not responsible for the observed `red/green' pattern. The latter might therefore be a true dynamical signature. The `red/green' transition does, however, not agree with the OAS. We can therefore state that there are no angle signatures consistent with $x_1(1)$ OLR orbits at the location of the classical \emph{fast bar}'s OLR.

This result of ruling out `Hercules' as an OLR candidate is in agreement with recent work that also looked at the extended phase-space structure of `Hercules' in \emph{Gaia} DR2 RVS. \citet{2020MNRAS.495..895B} found---using torus-based modeling---that the velocity structure of `Hercules', at different locations within $1~\text{kpc}$ from the Sun, is not consistent with trapping at the OLR. \citet{2019A&A...632A.107M} pointed out that the slope of the `Hercules' $\langle v_R \rangle$ feature in the $(L_z,\phi)$ plane, disagrees with the OLR prediction. Both these works prefer the CR as origin of `Hercules'.

\section{Discussion} \label{sec:discussion}

\subsection{Revisiting the OLR's orbit orientation flip in action space} \label{sec:OLR_flip_revisited}

Several strategies have been employed in the literature to identify the bar's OLR in the local stellar phase-space (e.g., \citealt{2000AJ....119..800D,2003A&A...401..975M,2007A&A...467..145C,2010MNRAS.407.2122M,2014AA...563A..60A,2017MNRAS.465.1443M,2017RAA....17..114T,2019A&A...626A..41M,2020A&A...634L...8K,2020ApJ...893..105H}; \citetalias{2021MNRAS.500.2645T}). Many use the well-known fact that the orientation of disk orbits with respect to the bar flip at the OLR from $x_1(2)$ orbits inside of the OLR to $x_1(1)$ orbits outside of the OLR (e.g., \citealt{1989A&ARv...1..261C,1991dodg.conf..323K,1993RPPh...56..173S,2000AJ....119..800D,2001A&A...373..511F,2008gady.book.....B,2019MNRAS.488.3324F}). At the Solar azimuth $\sim 25^\circ$ behind the bar, this translates to a flip from outward- to inward-moving stars.

In \citetalias{2021MNRAS.500.2645T}, we used this orbit orientation flip to point out a simple strategy to directly read off all possible OLR candidates and associated bar pattern speeds from the $(L_z,J_R)$ plane colour-coded by the radial velocity. Specifically, the `fraction of inward moving stars', $N_*(v_R < 0)/N_{*,\text{tot}}$, tells us which features in action space are predominantly outward-moving (colour-coded 
`red') or inward-moving (colour-coded `blue'). This is a proxy of the average radial phase, $\langle \theta_R \rangle$, rather than the average radial velocity, $\langle v_R \rangle$. We found that in simulations and for survey volumes out to several kpc, the OLR's $v_R$-signature can be studied very well in axisymmetric action space, where the ARL neatly separates the `red' from the `blue' feature.

In \citetalias{2021MNRAS.500.2645T}, we identified `red/blue' features around OLR ARLs in \emph{Gaia} DR2 RVS for the following pattern speeds:
\begin{itemize}[leftmargin=*,topsep=1ex,itemsep=1ex]
\item $\Omega_\text{bar} = 33~\text{km/s/kpc} = 1.2\Omega_0$, with the \texttt{Hat} as the inward-moving OLR ridge, and the gap between `Hat' and `Sirius' as the outward-moving part of the signature (c.f. the \emph{slow bar} in, e.g., \citealt{2019AA...626A..41M,2019MNRAS.490.1026H}),
\item $\Omega_\text{bar} = 45~\text{km/s/kpc} = 1.63\Omega_0$, with the outward-moving \texttt{Sirius} as the OLR ridge, and the `Hyades' as the inward-moving feature (c.f. \citealt{1991dodg.conf..323K}),
\item $\Omega_\text{bar} = 51~\text{km/s/kpc} = 1.85\Omega_0$, with \texttt{Hercules} as the inward-moving feature, and the `Horn' as the OLR ridge,  (c.f. the \emph{fast bar} in, e.g., \citealt{2000AJ....119..800D,2014AA...563A..60A}).
\end{itemize}

We refer the reader to \citetalias{2021MNRAS.500.2645T}, where we have discussed these candidates and their previous appearances in the literature in depth.

\subsection{Comparing the action- and angle-based methods for the OLR} \label{sec:action_vs_angle_based_methods}

There is more than one feature in the local kinematics that looks like the bar OLR's velocity flip. But only one can be the true OLR (if at all visible) and the others will be related to other bar resonances (e.g., \citealt{2019AA...626A..41M,2018MNRAS.477.3945H,2020MNRAS.499.2416A}) and spiral arms \citep{2004MNRAS.350..627D,2007A&A...467..145C,2018MNRAS.480.3132Q,2018ApJ...863L..37M,2019MNRAS.490.1026H,2020A&A...634L...8K,2020MNRAS.497..818H}. Our angle-based method provides additional constraining evidence.

In Figure \ref{fig:angle_space_candidate_OLR}, we compile for all bar pattern speed candidates the two different OLR diagnostics: (i) the `red/blue' feature in action-space around the OLR ARL from \citetalias{2021MNRAS.500.2645T} (summarized in Section \ref{sec:OLR_flip_revisited}), and (ii) the `slope of $-2$' feature in angle space from this work (see Section \ref{sec:OLR_candidates}). 

Interestingly, pattern speed candidates around $1.2\Omega_0$ and $\sim1.6\Omega_0$ were \emph{independently} suggested by both strategies. While for $1.2\Omega_0$, the OLR ARL falls right between a `red/blue' feature, this is not the case for our most promising candidate from angle space, $1.4\Omega_0$. For $1.4\Omega_0$, the ARL falls on top of the `blue', inward-moving \texttt{Sirius (G2, orange)} ridge, with a prominent outward-moving underdensity at \emph{higher} rather than smaller $L_z$ (c.f. \citetalias[\S 6.3.4]{2021MNRAS.500.2645T}). For $1.6\Omega_0$ or $1.74\Omega_0$, the ARL separates `red/blue' features only at small or high $J_R$, respectively. The best candidate from action and velocity space, the \texttt{Hercules} pattern speed $1.85\Omega_0$, was strongly disfavoured in Section \ref{sec:OLR_candidates} due to not exhibiting the expected OLR pattern in angle space.

\begin{figure*}
    \centering
    \includegraphics[width=\textwidth]{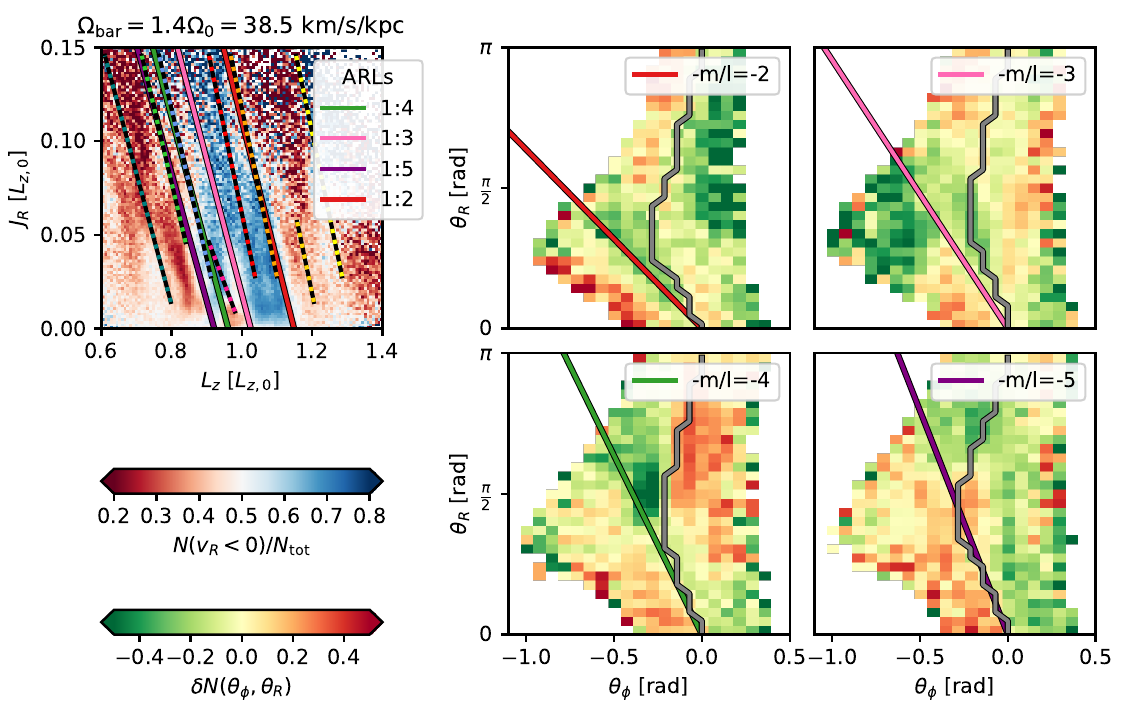}
    \caption{Angle space for stars in \emph{Gaia} DR2 at higher order resonances $(l = +1, m\in\{3,4,5\})$ for the pattern speed candidate of $\Omega_\text{bar}=1.4\Omega_0=38.5~\text{km/s/kpc}$ that exhibits an especially convincing OLR signature ($m=2$). Interestingly, $\Delta \theta_R/\Delta \theta_\phi = -m/l$ signatures are observed in these panels. For $m>2$, we expect these slopes resulting directly from the orbit shapes. Even though we could not reproduce these observations with our idealised simulated disk populations, the coincidence in the data is still noteworthy. The upper left panel shows the location of the ARLs (solid lines) with respect to the action ridges (dotted lines). The 1:5 and 1:4 ARLs align with sub-ridges in `Hercules', \texttt{Hercules (C1, green)} and \texttt{Hercules/Horn (D1, blue)}, respectively (c.f. \citetalias{2019MNRAS.484.3291T}).}
    \label{fig:higher_order_resonances}
\end{figure*}

The different outcomes of the two methods for a given OLR candidate can reveal additional information about the true dynamical nature of the respective feature. It is therefore important to point out that these two diagnostics differ from each other as follows: 
\begin{itemize}[leftmargin=*,topsep=1ex,itemsep=1ex]
    \item The `action-$v_R$-ARL' method requires both $x_1(1)$ \emph{and} $x_1(2)$ orbits, but checks only for the associated flip in $v_R$.
    \item The `angle-$\delta N$' method uses \emph{only} the stars on (possible) $x_1(1)$ orbits, but explicitly tests if the angle distribution is consistent with the expected \emph{shape} of these orbits.
\end{itemize} 

What mechanisms cause features that look like an OLR diagnostic in one method but not in the other? Coming up with alternative explanations is beyond the scope of this work. Considering the complex interplay of perturbers in the MW, dynamical studies of self-consistent and cosmological simulations might be required.

What is possible with our test particle simulation, however, is to bring together our action- and angle-based methods and derive two further idealized diagnostics for the bar OLR in angle space. In Appendix \ref{app:OLR_flip_x12}, we search for the orbit orientation flip in angle space, finding that we cannot find signatures of anti-aligned $x_1(2)$ orbits in the \emph{Gaia} data. In Appendix \ref{app:bar_strength_angle_space}, we discuss the strength of the OLR signature in angle space, which lets us rule out $\Omega_\text{bar} = 1.74\Omega_0 = 47.9~\text{km/s/kpc}$ from Section \ref{sec:OLR_candidates} as a candidate for the bar's pattern speed.

\subsection{Higher-order resonances: A curious coincidence in the \emph{Gaia} data for $\Omega_\text{bar}=1.4\Omega_0=38.5~\text{km/s/kpc}$} \label{sec:curious_coincidence}

In this work, we focused on identifying OLR $x_1(1)$ orbits, as those left the strongest imprints in our test particle simulations. We have also searched for $\Delta \theta_R / \Delta \theta_\phi = -m/l$ signatures of other bar resonances in the \emph{Gaia} angle data---but with less success. We provide some details in Appendices \ref{app:11_resonance} for the 1:1 resonance, in \ref{app:CR_resonance} for CR, and in \ref{app:14_resonance} for the 1:4 resonance.

Figure \ref{fig:higher_order_resonances}, however, shows an interesting observation we made in the angle data for the \texttt{MWPotential2014}. We selected stars in the \emph{Gaia} DR2 RVS action data next to the 1:4, 1:3, and 1:5 ARLs for our best $\Omega_\text{bar}$ candidate in Figure \ref{fig:angle_space_candidate_OLR}, namely $1.4\Omega_0 = 38.5~\text{km/s/kpc}$. In case of this 1:4 resonance---for which the ARL aligns with the \texttt{Hercules/Horn (D1, blue)} ridge and which contains also stars from the \texttt{Hyades/Horn (E1, pink)} ridge---we observe a `red/green' transition in angle space along a slope of -4. For the 1:3 resonance---which falls between the \texttt{Hyades/Horn (E1, pink)} and the \texttt{Sirius (F1, red)} ridges---a transition line with a -3 slope shows up for small $\theta_R$. For the 1:5 resonance---which aligns its ARL with the \texttt{Hercules (C1, green)} ridge and contains also stars from the \texttt{Hercules (C2, blue)} ridge---a transition line with slope -5 at small and larger $\theta_R$ is observed.

Based on our simulations, the `red/green' transition at this 1:4 resonance would suggest that the MW has a pointy bar with extremely strong $m=4$ component (see Appendix \ref{app:14_resonance}), which is not realistic. We can therefore not explain this observation on the basis of our simple simulations. Selection effects might also play a role, especially at the 1:4 and 1:5 resonance. Additional substructure is also present at $\theta_R\gtrsim\pi/4$ for the 1:3 resonance, and around $\theta_R \sim \pi/2$ for the 1:5 resonance. 

Partly responsible for this observed coincidence is also the choice of potential model. A model with a different rotation curve would not necessarily have the best OLR candidate and these higher-order resonances at the same pattern speed (see Appendix \ref{app:Eilers_pot}).

\citet{2019A&A...626A..41M} identified candidates for higher-order bar resonances in action-velocity space for a similar bar pattern speed. Because of the differences in the assumed potential model, their 1:4 resonance was related to \texttt{Sirius}, the 1:3 resonance to the small \texttt{gold (H)} ridge in the \emph{Gaia} data, and the \texttt{Horn/Hyades} to the 1:6 resonance. The assignment of ridges to resonances by \citet{2019A&A...626A..41M} therefore clearly differs from the one suggested here.

Interestingly, \citet{2020MNRAS.499.2416A} found higher-order bar resonance features in the self-consistent MW simulation by \citet{2019MNRAS.482.1983F} that might resemble the velocity structure of the `Hercules' sub-arches. Their `Hercules 2', `Hercules 1', and `Horn' arches would in this case be associated with 1:5, 1:4, and 1:3 bar resonances. Their `Hercules 2' and `Hercules 1' correspond to our \texttt{Hercules (C1, green)} and \texttt{Hercules/Horn (D1, blue)} ridges. The conclusion by \citet{2020MNRAS.499.2416A}, even though based on a completely different method, is therefore consistent with our findings.

\citet{2021MNRAS.508..728K} argue based on the action ridges in \emph{Gaia} EDR3 that both these proposals for higher-order resonances are plausible, also seeing indications for additional -1:4, +3:4, and +1:1 bar resonances. 

\subsection{Discussion of previous work on resonant angle signatures at the `Hyades'} \label{sec:discussion_Hyades_papers}

In the past, the observational restriction to stars within $\sim 200~\text{pc}$ prohibited a more extensive exploitation of the phase-angles. Stars with $L_z \ll L_{z,0}$ were only observed close to apocenter, those with $L_z \gg L_{z,0}$ close to pericenter. Only for orbits with $L_z \sim L_{z,0}$ could a large range of radial phases be probed. The `Hyades' moving group, $(U, V) \sim (-33, -16)~\text{km/s}$, and the associated ridge in action space around $L_z\sim 0.95L_{z,0}$, was therefore the best candidate for an action-angle exploration.

\citet{2010MNRAS.409..145S} was the first to use angle space as a diagnostic for resonances. When investigating action space, he found that the ridge of the `Hyades' stream lies along a line on which the resonance condition \eqref{eq:axisym_res_cond} is satisfied, i.e. along an ARL in our nomenclature. ARLs for different $m$ and $l$ have, however, quite similar slopes, and scattering at the resonances can also affect the orientation of the ridge with respect to the ARL in action space. \citet{2010MNRAS.409..145S} therefore proposed to use angle space to constrain the $m$ and $l$ of this resonance. In particular, he suggested to use the relation
\begin{equation}
    m \theta_\phi + l \theta_R \simeq \text{const}. \Longrightarrow \frac{\Delta \theta_\phi}{\Delta \theta_R} = - \frac ml \label{eq:sellwood_angle_relation}
\end{equation}
for stars trapped at a resonance (see Section \ref{sec:background_slow_angle_theory}). From analysing data from the Geneva-Copenhagen Survey within $200~\text{pc}$ (GCS; \citealt{2009A&A...501..941H}), he concluded that stars in the \texttt{Hyades} ridge lie along $\Delta \theta_\phi/\Delta \theta_R > 0$, being indicative of an inner Lindblad resonance of a slowly rotating spiral arm.

However, \citet{2011MNRAS.418.1565M} demonstrated that this conclusion was mainly driven by selection effects: Firstly, the asymmetric drift in the Solar neighbourhood lead to a general excess of stars close to apocenter. Secondly, the limited survey volume restricted the observable region in angle plane predominantly to the two diagonally opposite quadrants $(\text{`outward-moving'} \land \theta_\phi < 0)$ and $(\text{`inward-moving'} \land \theta_\phi > 0)$.\footnote{$\theta_R$ is periodic in $2\pi$, and usually presented as $\theta_R\in [-\pi,\pi]$. \citet{2010MNRAS.409..145S}, \citet{2011MNRAS.418.1565M}, and \citet{2011MNRAS.418.2459H} define $\theta_R = 0$ at apocenter and $\theta_R < 0$ is therefore 
`outward-moving'. We use the definition from \texttt{galpy}, where $\theta_R = 0$ at pericenter, and $\theta_R > 0$ denotes therefore 
`outward-moving'. This zero-point shift leaves Equation \eqref{eq:sellwood_angle_relation} unchanged.} Together, these biases lead to measurements of $\Delta \theta_\phi/\Delta \theta_R > 0$ for the `Hyades' stars. 

\citet{2011MNRAS.418.2459H} revisited the analysis using a different set of stars within 200pc, taken from the RAVE and SDSS surveys \citep{2008AJ....136..421Z,2009ApJS..182..543A}, confirming the claim by  \citet{2010MNRAS.409..145S} that the `Hyades' look like a resonant ridge in action space, and the conclusion by \citet{2011MNRAS.418.1565M} that the severe selection effects do not allow to distinguish between outer and inner Lindblad resonance and to determine $m$.

These authors have selected their candidates for stars scattered at a Lindblad resonance along the ARL in action space. This makes sense for an inner Lindblad resonance, given the alignment of the scattering direction with the ARL (c.f. fig. 5 in \citealt{2010MNRAS.409..145S}). For the OLR, we expect the resonant $x_1(1)$ orbit stars to be scattered on average off the ARL, towards higher $L_z$, while the $x_1(2)$ orbits lie on the lower-$L_z$ side of the ARL. In this work, where we have looked only at the OLR with $m=2$, we have explicitly made this distinction between the two orbit classes and used it as an additional diagnostic to identify the OLR in Appendix \ref{app:OLR_flip_x12}.

These authors have also stated that it is because of the smallness of the $200~\text{pc}$ Solar neighbourhood that it is difficult to determine the exact $m$ of the resonance based on a search for $\Delta \theta_R / \Delta \theta_\phi = - m/l$ features. In this work, thanks to the $3~\text{kpc}$ extent of \emph{Gaia} DR2, we were able to search specifically for the OLR with $m=2$. Based on our test particle simulations, we found that this OLR is actually the resonance for which this strategy can work best.

Concerning the `Hyades'---the main interest of \citet{2010MNRAS.409..145S}, \citet{2011MNRAS.418.1565M,2013MNRAS.430.3276M}, and \citet{2011MNRAS.418.2459H}---we first identified that it could be related to the bar's OLR based on the slope in angle space for $\Omega_\text{bar} = 1.74\Omega_0$ (in Section \ref{sec:OLR_candidates} and Figure \ref{fig:angle_space_scan_OLR}), but given the evolution of this signature that we expect with $L_z$ (or assumed $\Omega_\text{bar}$; in Appendices \ref{app:OLR_flip_x12}- \ref{app:bar_strength_angle_space}, Figure \ref{fig:angle_space_model_OLR}), we rule this candidate out. In Section \ref{sec:curious_coincidence} and Figure \ref{fig:higher_order_resonances}, we speculated that the `Hyades' could be related to the bar's 1:4 resonance. The work by \citet{2019AA...626A..41M} suggests the `Hyades/Horn' to be close to the bar's 1:6 resonance. In the past, some other authors \citep{2005AJ....130..576Q,2011MNRAS.415.1138P} have suggested that the `Hyades' could be caused by the -1:4 inner Lindblad resonance of a spiral arm. Angle space in the corresponding panels in Figure \ref{fig:angle_space_scan_OLR} does not obviously exhibit a feature of slope of $-m/l=+4$ or $-m/l=-6$ clearly distinct from the `$\langle \phi \rangle\sim0$' line (which itself exhibits slopes of -4 and +6), so we can neither support nor contradict this. The resonant origin of the `Hyades' remains therefore an open question.

\subsection{Discussion of previous work on angle asymmetries in \emph{Gaia} DR2}

Our study is based on comparing stellar numbers at positive and negative $(\theta_\phi,\theta_R)$ within $\sim L_z$ bins to uncover the characteristic orbit shape at the OLR. A recent study by \citet{2020ApJ...899L..14H} compared in the Gaia DR2 data stellar numbers at positive and negative Galactic azimuth $\phi$ within $R$ bins. They searched for a number asymmetry flip at $R_\text{OLR}$ induced along $\phi$ by the OLR's orbit shape and orientation flip (c.f., e.g., Figure A1 in \citetalias{2021MNRAS.500.2645T}). Because of the relations between $\theta_\phi \longleftrightarrow \phi$ and $L_z \longleftrightarrow R$, respectively, the basic idea in these two works is very similar.

\citet{2020ApJ...899L..14H} found $\Omega_\text{bar} = 49.3 \pm 2.2~\text{km/s/kpc} = 1.75\pm0.08 \Omega_0$ (for $\Omega_0 = (229.0~\text{km/s})/(8.122~\text{kpc})$; \citealt{2019ApJ...871..120E}) as their best candidate. This is consistent with the classic \emph{fast} \texttt{Hercules} \emph{bar} and our weak candidate from angle space with $\Omega_\text{bar} = 1.74-1.77\Omega_0$ (Figures \ref{fig:angle_space_candidate_OLR} and \ref{fig:angle_space_scan_OLR_Eilerspot}). 

Our study included the radial phase $\theta_R$ as an additional piece of information in the analysis. This changed the outcome of the asymmetry study significantly: It ruled out the \emph{fast} \texttt{Hercules} \emph{bar}, and introduced the OLR of a bar with pattern speed $1.4\Omega_0$ as a strong candidate, even though it misses the classic OLR orbit orientation flip.

\section{Summary and conclusion} \label{sec:summary}

In this work, we propose an independent strategy to identify the Outer Lindblad Resonance (OLR; $l=+1, m=2$) of the Galactic bar in the \emph{Gaia} DR2 RVS data within $3~\text{kpc}$, using the space of orbital angles $(\theta_\phi,\theta_R)$ (Figure \ref{fig:angles_in_epicycle_approximation_explained}) and the relation 
\begin{equation}
    \theta_R = -2 (\theta_\phi - \phi_\text{bar}) + \pi,
\end{equation}
(with slope $\Delta \theta_R / \Delta \theta_\phi = -m/l = -2$) characteristic for resonant orbits of the $x_1(1)$ OLR orbit class aligned with the bar (Figures \ref{fig:example_trapped_orbits}-\ref{fig:OLR_resonant_stars}).

Based on idealized test particle simulations within a barred galaxy potential, we illustrate that this diagnostic property can be made visible for the following sub-selection of stars: For the true pattern speed $\Omega_\text{bar}$, $x_1(1)$ orbits dominate the OLR's resonant scattering ridge in action space $(L_z,J_R)$ (on the high-$L_z$ side of the resonance line; Figures \ref{fig:fraction_of_resonant_stars}-\ref{fig:angles_selection_effects_explained}).

The completeness of surveys decreases with distance from the Sun. This effectively hides the resonant angle signatures. We show that by investigating the relative number difference $\delta N(\theta_\phi,\theta_R)$ of stars at $(\theta_\phi,\theta_R)$ vs. $(-\theta_\phi,-\theta_R)$, we can mitigate this selection effect and reveal the OLR's angle slope, as $\delta N=0$ along the line $\theta_R = -2 \theta_\phi$ (Figure \ref{fig:angle_mirroring_explained}).

This strategy can be affected if the survey's completeness also exhibits differences at Galactic longitudes $\pm l$ (i.e. at $+|l|$ vs. $-|l|$). The OLR signature should still be visible for completeness differences of less than 30\%, as is the case in the \emph{Gaia} DR2 RVS data. We find that marking the line along which the mean Galactic azimuth $\langle \phi \rangle \sim 0$ in the $(\theta_\phi,\theta_R)$ plane is a good diagnostic to gauge if the patterns in $\delta N$ in the angle plane could be affected by selection effects (Figure \ref{fig:issues_with_incompleteness}).

We present the \emph{Gaia} DR2 RVS angle data out to 3 kpc as a function of $L_z$---or similarly, as a function of the location of the OLR in action space (Figure \ref{fig:angle_space_scan_OLR}). This reveals phase-substructure that might be informative about the physical mechanisms causing the features at this location in action space.

By comparing this to the expected OLR $x_1(1)$ signature ($\delta N=0$ along $\theta_R = -2 \theta_\phi$), we extract several candidates for the MW's true OLR (Figure \ref{fig:angle_space_candidate_OLR}):

$\Omega_\text{bar}\sim1.2-1.27\Omega_0\sim 33-36~\text{km/s/kpc}$ explains the `Hat' moving group with the OLR of a \emph{slow bar}. It is consistent with a flip from outward-to-inward movements, classically expected at the OLR. However, the outward-moving part does not exhibit the angle signatures expected for the associated $x_1(2)$ orbits, anti-aligned with the bar (Figure \ref{fig:angle_space_model_OLR}). We note that the selection effects around this OLR radius ($R_\text{OLR} \sim 10.5-10.7~\text{kpc}$) are quite strong, affecting our conclusions. A larger coverage of the Galactic disk with precise phase-space data should resolve this. This pattern speed is a favourite in the literature as it associates the `Hercules' stream with the bar's CR resonance.

$\Omega_\text{bar}\sim1.4-1.45\Omega_0\sim 39-41~\text{km/s/kpc}$ is the most convincing candidate in angle space. It would associate part of the `Sirius' moving group with the OLR of a \emph{slightly faster slow bar} and agrees interestingly with measurements from the Galactic center. For this pattern speed, the outward-moving part of the OLR signature with $x_1(2)$ orbits is missing, however. If this was the true OLR, we require a dynamical mechanism that could explain the absence of $x_1(2)$ OLR orbits.

The angle planes at $\Omega_\text{bar}\sim1.6-1.65\Omega_0\sim 44-47~\text{km/s/kpc}$ and $\Omega_\text{bar}\sim1.74-1.77\Omega_0\sim 48-50~\text{km/s/kpc}$ also show signatures reminiscent of the OLR $x_1(1)$ orbits, with the resonance line falling between `Hyades' and `Sirius', and close to the `Hyades' and `Horn', respectively. But when investigating the orbital phase-structure in an extended $L_z$ region (or equivalently $\Omega_\text{bar}$-OLR range in action space) around it, angle space does not agree with our expectation (c.f. Figure \ref{fig:angle_space_model_OLR}).

In addition, we show that at the OLR of the classic \emph{fast bar} pattern speed, $\Omega_\text{bar}\sim1.85\Omega_0\sim 51~\text{km/s/kpc}$, which associates the `Horn' with the $x_1(1)$ and `Hercules' with $x_1(2)$ OLR orbits, angle space does not exhibit the OLR signature at all. This result is also robust against the influence of selection effects. We propose therefore that the \emph{Gaia} DR2 angle data rules out that the MW has a \emph{fast bar} (Figure \ref{fig:angle_space_candidate_OLR}).

We uncovered a curious coincidence in angle space related to the higher-order resonances of the strong bar candidate with $\Omega_\text{bar}=1.4\Omega_0$, the \emph{slightly faster slow bar}. Close to its $l$:$m$ resonances with $l=+1$ and $m\in \{3,4,5\}$, the angle space of \emph{Gaia} DR2 exhibits features with $\delta N=0$ along lines $\theta_R = -m/l \cdot \theta_\phi$ (Figure \ref{fig:higher_order_resonances}). On the one hand, this could further support this pattern speed candidate, as we expect this slope in theory from the intrinsic orbit shapes. On the other hand, we could not explain this observation based on our test particle simulations. The true nature of these features remains an open question.

\emph{Gaia} DR3 will provide accurate 6D phase-space measurements for stars even beyond $3~\text{kpc}$. For this larger coverage of the Galactic disk, our proposed strategy to search for resonance signatures of the central bar in the space of orbital phase angles promises new insights into the dynamical mechanisms that have shaped our Galaxy.

\section*{Data availability}

This work has made use of data from \emph{Gaia} DR2 \citep{2016A&A...595A...1G,2018A&A...616A...1G} available at \url{https://gea.esac.esa.int/archive}. For \emph{Gaia} DR2's radial velocity sample \citep{2019A&A...622A.205K}, stellar distances were taken from \citet{2019MNRAS.487.3568S} and are available at \url{https://zenodo.org/record/2557803}. 

Action-angle estimation and test particle simulations underlying this article were produced with the \texttt{galpy} code by \citet{2015ApJS..216...29B} which is publicly available at \url{http://github.com/jobovy/galpy}. The action-angle data and simulations will be shared on reasonable request to the corresponding author.

\section*{Acknowledgements}

The author thanks Jan Rybizki for providing the \emph{Gaia} DR2 RVS completeness function prior to publication, Francesca Fragkoudi for sharing code to derive OLR parent orbits, Rebekka Bieri and Simon White for valuable comments on a first paper draft, and the MPA's Dynamics group for helpful discussions. The author thanks also the referee James Binney for instructive suggestions to improve especially the theory part of this paper. 

This work has made use of data from the European Space Agency (ESA) mission \emph{Gaia} (\url{https://www.cosmos.esa.int/Gaia}), processed by the \emph{Gaia} Data Processing and Analysis Consortium (DPAC, \url{https://www.cosmos.esa.int/web/Gaia/dpac/consortium}). Funding for the DPAC has been provided by national institutions, in particular the institutions participating in the \emph{Gaia} Multilateral Agreement.

This project was developed in part at the 2019 Santa Barbara Gaia Sprint, hosted by the Kavli Institute for Theoretical Physics at the University of California, Santa Barbara.

This research was supported in part at KITP by the Heising-Simons Foundation and the National Science Foundation under Grant No. NSF PHY-1748958.

%%%%%%%%%%%%%%%%%%%% REFERENCES %%%%%%%%%%%%%%%%%%

% The best way to enter references is to use BibTeX:
\bibliographystyle{mnras}
\bibliography{bibliography} 

%%%%%%%%%%%%%%%%%%%%%%%%%%%%%%%%%%%%%%%%%%%%%%%%%%

\appendix

\section{Illustrating parent, librating, and circulating orbits around the OLR} \label{app:slow_angle_frequencies}

\begin{figure*}
    \centering
    \includegraphics[width=0.68\textwidth]{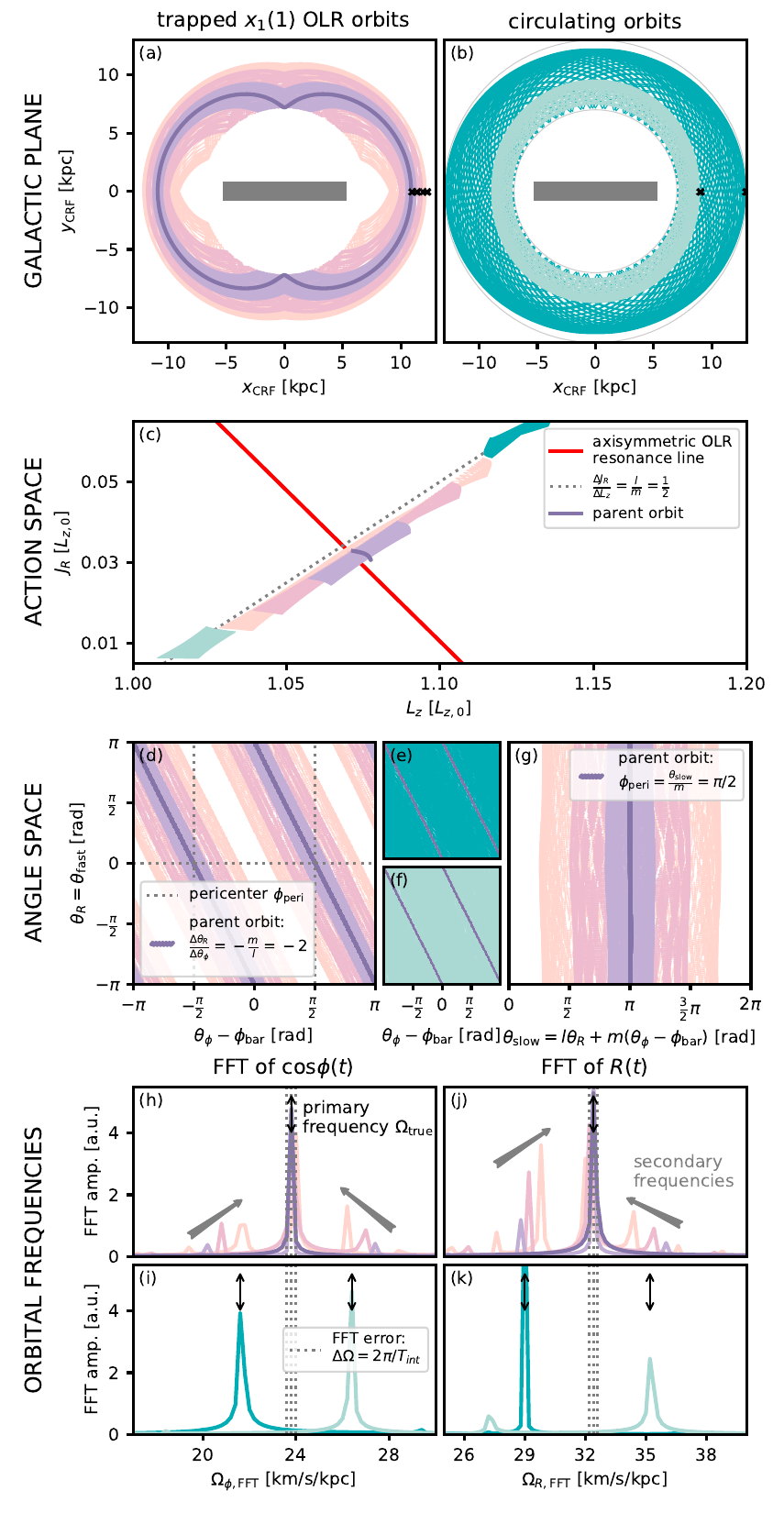}
    \caption{Comparing trapped resonant orbits at the OLR with circulating orbits in the Galactic disk plane co-rotating with the bar, as well as in action-angle-frequency space. For more details see Section \ref{sec:background_slow_angle_theory} and Appendix \ref{app:slow_angle_frequencies}.}
    \label{fig:slow_angle_frequencies}
\end{figure*}

In Section \ref{sec:theory}, we summarized the behaviour of resonant orbits in a barred Galaxy potential in action-angle space. Figure \ref{fig:slow_angle_frequencies} illustrates this now further with a few example orbits close to the OLR. We use the bar potential from the \texttt{Fiducial\_40} test particle simulation (Section \ref{sec:simulation_description}) with pattern speed $\Omega_\text{bar} = 40~\text{km/s/kpc}$.

\subsection{Resonant orbits}

Based on the algorithm by \citet{2015MNRAS.450..229F} (see also \citealt{1992MNRAS.259..328A,1993RPPh...56..173S}), we selected a parent orbit at the OLR with $l=+1$ and $m=2$ from the $x_1(1)$ orbit class, which is shown in purple in Figure \ref{fig:slow_angle_frequencies}. Panels (h) and (j) show the fast Fourier transform (FFT) spectra of the parent orbit. Each has one dominant frequency, $\Omega_{R,\text{true}}$ and $\Omega_{\phi,\text{true}}$, in $\Omega_{R,\text{FFT}}$ and $\Omega_{\phi,\text{FFT}}$, respectively, which satisfy the resonance condition in Equation \eqref{eq:general_res_cond}. For the parent orbit, $\Omega_\text{$i$,true} \approx \Omega_\text{$i$,axi}$, i.e. it lives almost exactly on top of the ARL (Equation \ref{eq:axisym_res_cond}) in action space (Panel (c)). This example parent orbit is launched from a position with initial radius $R_\text{parent} = 10.89~\text{kpc}$ along the $x_\text{CRF}$ axis, its Jacobi energy is $E_\text{Jacobi} = -112,553~\text{km}^2/\text{s}^2$ and $J_z\approx0$. Different choices of $E_\text{Jacobi}$ would position the closed periodic orbits higher or lower on the ARL and at larger or smaller initial $R$ (see \citetalias[fig. 3]{2021MNRAS.500.2645T}). Larger $J_z$ would move the parent orbit together with the ARL to lower $L_z$ (see \citetalias[\S5.2.2]{2021MNRAS.500.2645T}).

All orbits in this figure have the same $E_\text{Jacobi}$, but slightly shifted initial position $R = f \times R_\text{parent}$ (black crosses in Panels (a) and (b)).

We show librating resonant OLR orbits in lighter shades of pink and purple in Figure \ref{fig:slow_angle_frequencies}. We used  $f \in 1+\left\{0.04,0.09,0.12 \right\}$ to initialize them; orbits with $f \in 1-\left\{0.04,0.09,0.12 \right\}$ would look almost identical. They librate around their parent orbit with successively larger amplitude in actions, angles, and pericenter. The grey dotted line in Panel (c) denotes a line of constant fast action. The primary frequencies of the librating orbits are the same as those of the parent orbit, i.e. they satisfy Equation \eqref{eq:general_res_cond} exactly. As opposed to the parent orbit, their FFT spectra exhibit also secondary frequencies which are related to the libration (Panels (h) and (j); see also \citealt{1982ApJ...252..308B}).

\subsection{Circulating orbits}

The orbits in shades of green in Figure \ref{fig:slow_angle_frequencies} were initialized with larger $f \in \left\{1.185,0.82\right\}$. Their pericenters are not trapped anymore in $\phi_\text{CRF}$, but rather \emph{circulating} around the bar. This can also be seen in Panels (e) and (f) for the angle coordinate $\theta_\phi$ with respect to the bar. The dark green orbit lives outside of the OLR and its symmetry is aligned with the bar; the light green orbit lives inside of the OLR and its symmetry is anti-aligned with the bar. We are seeing here the orbit orientation flip at the OLR. Also in action space, Panel (c), the circulating orbits are not trapped around the ARL anymore and truly live at $L_z$s inside or outside of the resonance. They have clear primary frequencies $\Omega_\text{$i$,true}$ and no substantial secondary frequencies. The primary frequency `jumps' from the librating orbits with the exact resonance frequencies to the circulating orbits whose $\Omega_{i,\text{true}}$ are close to the former's secondary libration frequencies \citep{1982ApJ...252..308B}. The trapping of the resonant orbits therefore manifests also in frequency space.

\emph{Special cases.} For close-to-circular orbits $J_R \sim 0$ \citep{2020MNRAS.495..886B} and in the small transition region from librating and to circulating orbits, the behaviour is more complex (e.g. \citealt{1982ApJ...252..308B}), but beyond the scope of this paper.

\section{The full angle space of \emph{Gaia} DR2 RVS: Searching for the OLR} \label{app:Gaias_full_angle_space_MWPot}

\begin{figure*}
    \centering
    \includegraphics[width=0.9\textwidth]{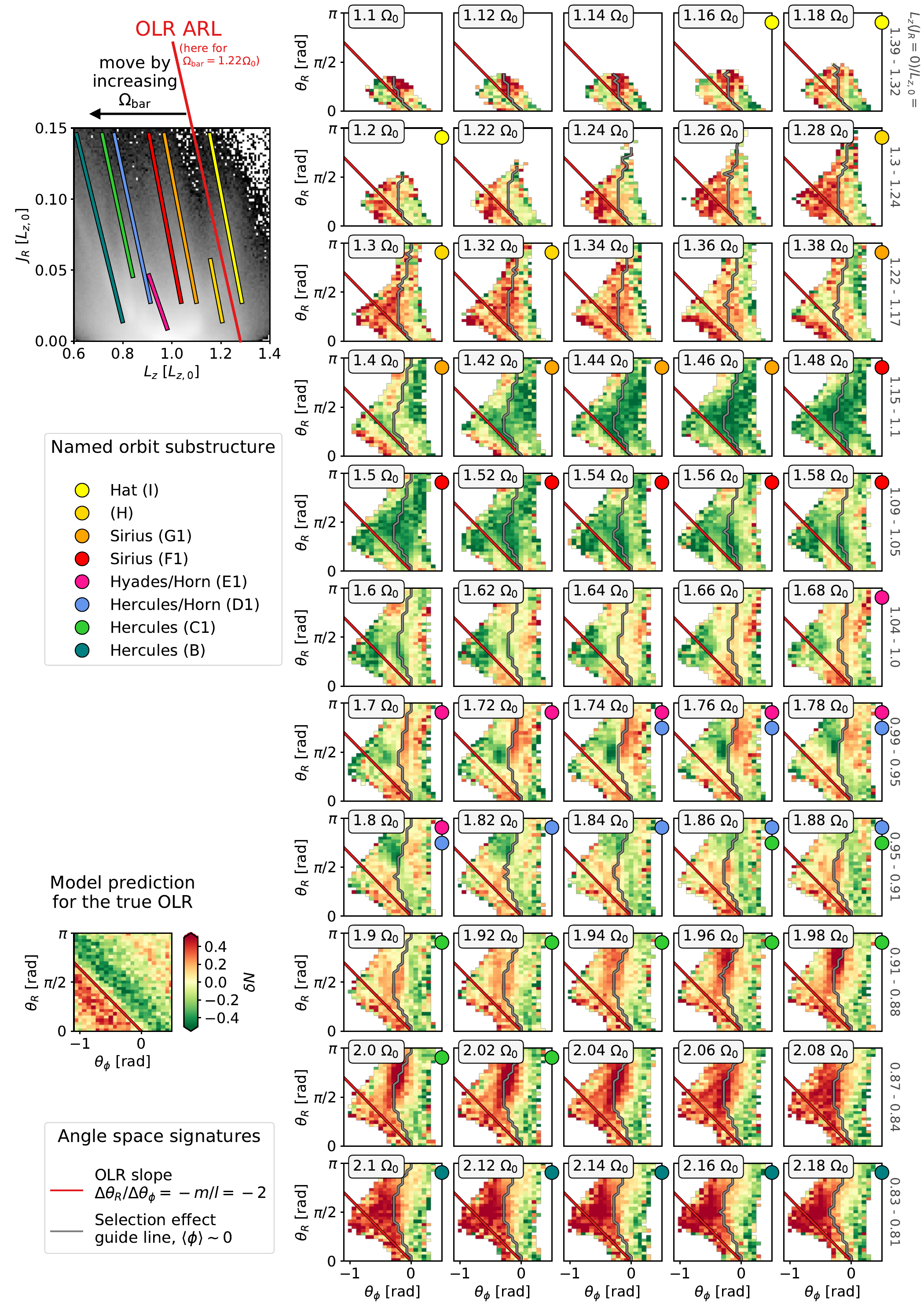}
    \caption{Searching for OLR candidates in the \emph{Gaia} DR2 RVS angle data. Varying the bar pattern speed $\Omega_\text{bar}$ moves the OLR ARL across the action plane (\emph{upper left panel}). We show the angle space asymmetry $\delta N$ for stars next to the OLR ARL for different $\Omega_\text{bar}$ (\emph{right panels}). We compare the $\delta N$ in each panel to the prediction from the simulation (\emph{lower left panel}). For a detailed discussion of the most promising OLR candidates, see Figure \ref{fig:angle_space_candidate_OLR}.}
    \label{fig:angle_space_scan_OLR}
\end{figure*}

In Figure \ref{fig:angle_space_scan_OLR}, we present the full 4D action-angle data of the \emph{Gaia} DR2 RVS sample.

In the upper left panel, we mark the action overdensity ridges, i.e., the orbit substructures associated with the well-known moving groups in the Solar neighbourhood \citepalias{2019MNRAS.484.3291T}.
Each 2D bin in action space has a different 2D angle distribution. We reduce this complexity to 3D, by binning the $(L_z,J_R)$ plane into slices roughly parallel to the action ridges. The slices are parameterized in terms of the OLR ARL position of an assumed $\Omega_\text{bar}$ (see Section \ref{sec:OLR_candidates} for details).

Each angle space panel in Figure \ref{fig:angle_space_scan_OLR} contains the stars from one of these actions slices. We show the stellar number asymmetry $\delta N(\theta_\phi,\theta_R)$ (Equation \ref{eq:delta_N}) to make substructure better visible (see Section \ref{sec:selection_effects_with_distance}). Overplotted in grey is the `$\langle \phi \rangle \sim 0$' line that indicates that some features might still be related to selection effects (see Section \ref{sec:selection_effects_with_longitude}).

To help with the orientation in \emph{Gaia}'s angle space---which angle panel belongs to which region in action space---we provide in Figure \ref{fig:angle_space_scan_OLR}:
\begin{enumerate}[leftmargin=*,topsep=1ex,label=(\roman*)]
    \item for each row of angle panels on the very right the rough $L_z$-range (at $J_R \sim 0$) that stars in these slices of action-angle space have.
    \item for each angle panel the assumed $\Omega_\text{bar}$ in units of $\Omega_0$ for which the respective action slice would fall together with the location of the OLR ridge.
    \item for some angle panels a coloured dot in the upper right, which marks that this slice contains stars from one of the famous orbit substructures.
\end{enumerate}

For example, in the panels marked by the pink dot, standing for the \texttt{Hyades/Horn (E1, pink)} ridge, we see an outward-moving (`red') clump at $\theta_R\sim 0.6\pi$, which contains stars from the `Hyades' moving group. The inward-moving (`green') clump at $\theta_R\sim \pi/2$ in the same panels is the large scale analogue of the `Horn' feature (around $(U,V)\sim(60,-25)
~\text{km/s}$) in local velocity space.

As laid out in Section \ref{sec:OLR_candidates}, Figure \ref{fig:angle_space_scan_OLR} provides the basis to search for candidates for the MW's true bar OLR.

\section{Further constraints on bar resonances from angle space}

\begin{figure*}
    \centering
    \includegraphics[width=\textwidth]{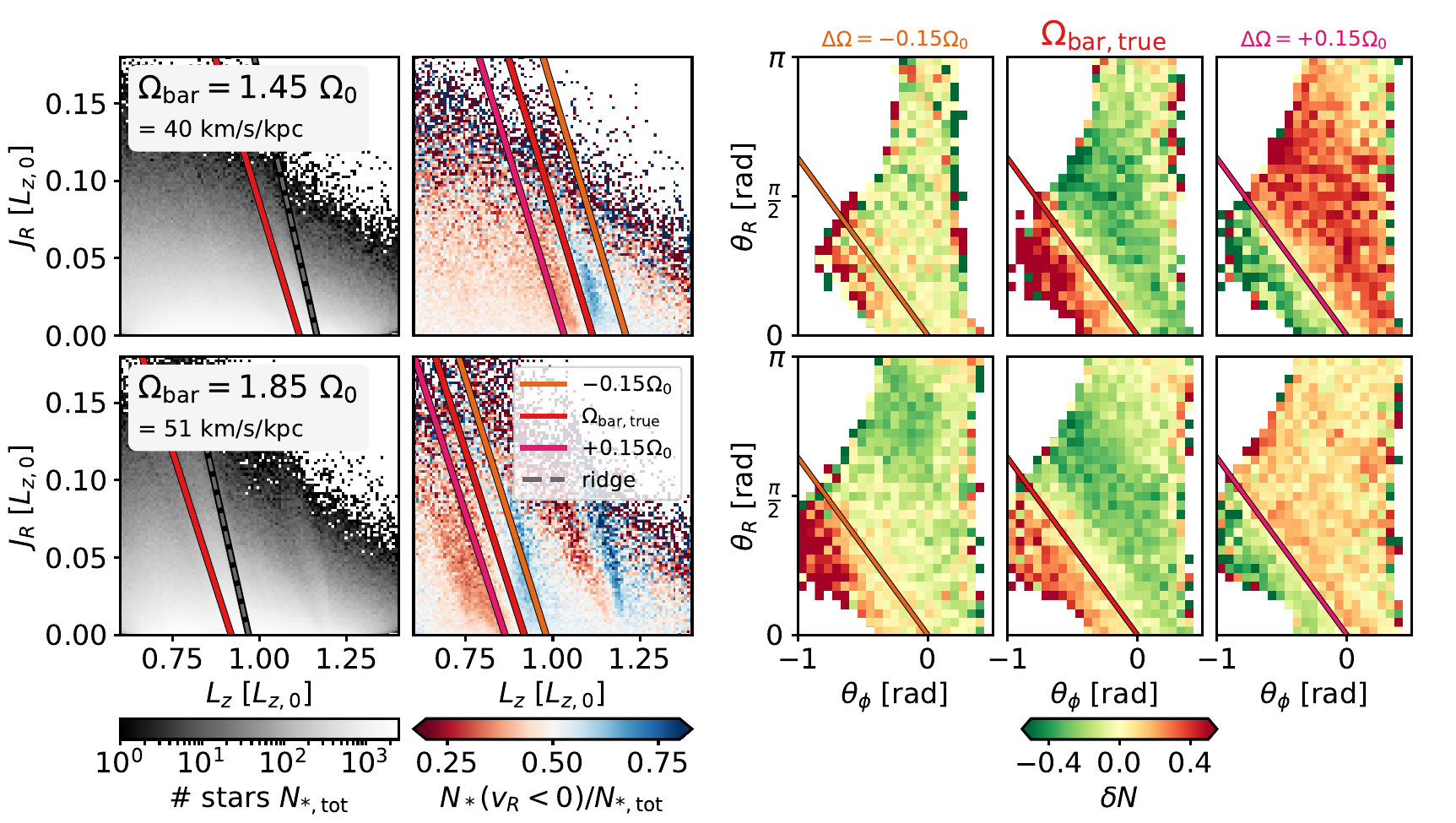}
    \caption{The orbit orientation flip at the OLR---causing the `red/blue' feature in action space (\emph{2nd column})---can be seen in the \texttt{Fiducial\_40} and \texttt{Fiducial\_51} simulations also in angle space (\emph{4th vs. 5th column}). No such angle asymmetry flip is observed in the \emph{Gaia} angle data in Figure \ref{fig:angle_space_scan_OLR}. Apart from this, the simulations tell us, that the OLR angle slope of $-2$ should be visible over a range of assumed $\Omega_\text{bar}$ (\emph{3rd column}). This supports the \emph{slow} \texttt{Hat} or \emph{slightly faster slow} \texttt{S19B19} $\Omega_\text{bar}$.}
    \label{fig:angle_space_model_OLR}
\end{figure*}

The OLR is the strongest signature of the bar outside of CR. In this work, we focused therefore on finding the signature of resonant $x_1(1)$ OLR orbits in angle space. Based on our idealized test particle simulations, the angles can be used as a diagnostic even beyond that. In the following, we will therefore shortly discuss (i) anti-aligned $x_1(2)$ orbits close to the OLR (Section \ref{app:OLR_flip_x12}), and (ii) the strength of the OLR signature (Section \ref{app:bar_strength_angle_space}), (iii) 1:1 resonant orbits (Section \ref{app:11_resonance}), (iv) CR orbits (Section \ref{app:CR_resonance}), and (iv) 1:4 resonant orbits (Section \ref{app:14_resonance}).

\subsection{The OLR's orbit orientation flip in angle space} \label{app:OLR_flip_x12}

Following the discussion in Sections \ref{sec:OLR_flip_revisited}-\ref{sec:action_vs_angle_based_methods}, we expect that the OLR's classic orbit orientation flip should not only show up in action-$v_R$ space, but also in angle-$\delta N$ space. This provides an additional diagnostic for the bar's OLR.

In Figure \ref{fig:angle_space_model_OLR}, right-most column, we show for the \texttt{Fiducial\_40} and \texttt{Fiducial\_51} simulations angle space for stars that were taken to the left of the true OLR ridge from the outward-moving, underdense part of the OLR signature. (We do this by moving the OLR ARL by $\Delta \Omega_\text{bar}=+0.15\Omega_0$ with respect to the true $\Omega_\text{bar}$.) This selection contains many stars on circulating orbits that are elongated in an \emph{anti-aligned} fashion with respect to the bar with a 1:2 symmetry and belong to the $x_1(2)$ orbit class. The corresponding angle plane exhibits a `green/red' pattern with OAS $-2$, just as the $x_1(1)$ orbits in the OLR ridge (4th column in Figure \ref{fig:angle_space_model_OLR}), but with the colours reversed.

Consequently, we now search in the \emph{Gaia} data in Figure \ref{fig:angle_space_scan_OLR} for a panel that is `green' \emph{below} the OAS line and `red' \emph{above} it.

The $1.1-1.14\Omega_0$ panels are the only candidate. However, this is on the wrong side of the \texttt{Hat} bar's OLR ARL---i.e. $\Delta \Omega < 0$ instead of the required $\Delta \Omega > 0$---to support the candidate $\Omega_\text{bar}=1.2\Omega_0$. This part of phase-space is also strongly affected by selection effects.

The $1.46-1.58\Omega_0$ panels do have $\Delta \Omega >0$ with respect to our strong candidate $\Omega_\text{bar}=1.4\Omega_0$. But while there is `green' below the OAS line, there is no `red' feature above the line. This is the angle equivalent of the missing outward-moving `red' feature in action-$v_R$ space for this pattern speed from \citetalias[\S6.3.4]{2021MNRAS.500.2645T} and also shown in Figure \ref{fig:angle_space_candidate_OLR}.

In conclusion, we do not find any signatures of $x_1(2)$ OLR orbits in the \emph{Gaia} angle space.

\subsection{Strength of the OLR signature in angle space} \label{app:bar_strength_angle_space}

The strength of the bar determines how large the region in action space is within which stars are on trapped OLR $x_1(1)$ orbits. This region is bounded by the separatrix and determines the width of the action-space scattering ridge. The OAS signature should be visible across the whole ridge (c.f. Figure \ref{fig:fraction_of_resonant_stars}).

A shift in the assumed $\Omega_\text{bar}$ of $\Delta \Omega =+0.02\Omega_0$, as used for the \emph{Gaia} angle exploration in Figure \ref{fig:angle_space_scan_OLR}, corresponds in the OLR ARL to a shift of $\Delta L_z \sim -0.01L_{z,0}$.\footnote{The shift $\Delta \Omega =+0.02\Omega_0$ in $\Omega_\text{bar}$ corresponds to slightly different shifts $\Delta L_z$ of the OLR ARL for different pattern speeds: for example, $\Delta L_z = -0.017$ or $-0.008L_{z,0}$ at $\Omega_\text{bar}=1.2$ or $1.85\Omega_0$, respectively.} This is smaller than the width of the OLR ridge we observe in the simulations, $|\Delta L_z| \sim 0.05L_{z,0}$. We therefore expect to see the OLR angle signature in a range of panels in Figure \ref{fig:angle_space_scan_OLR}.

In the simulations, we even observed the OLR signature over ranges of
\begin{itemize}[leftmargin=*,topsep=1ex,itemsep=1ex]
    \item $\Omega_\text{bar} = {\Omega_\text{bar,true}}^{+0.02}_{-0.1}$ (\texttt{Fiducial\_40}) and
    \item $\Omega_\text{bar} = {\Omega_\text{bar,true}}^{+0.01}_{-0.2}$ (\texttt{Fiducial\_51}).
\end{itemize}
The range in the \texttt{Fiducial\_51} simulation is larger, because we chose the bars in both simulations to have the same strength at $R_0=8~\text{kpc}$. Consequently, the bar influence is stronger at $R_\text{OLR}(51~\text{km/s/kpc})=7.3
~\text{kpc}$ than at $R_\text{OLR}(40~\text{km/s/kpc})=9~\text{kpc}$. To illustrate this, we show in the 3rd column of Figure \ref{fig:angle_space_model_OLR} the angle planes for $\Delta \Omega = -0.15\Omega_0\sim-4~\text{km/s/kpc}$ for both simulations. The stronger the bar at $R_\text{OLR}$, the more pronounced the scattering ridge, the larger the assumed $\Omega_\text{bar}$ range over which the OLR signature should show up in angle space.

This has two implications.

Firstly, this introduces an intrinsic uncertainty in our angle-based method to measure the bar's pattern speed. For the pattern speed candidates identified in Section \ref{sec:OLR_candidates}, we adopt based on Figure \ref{fig:angle_space_scan_OLR} the following rough ranges of:
\begin{eqnarray}
\Omega_\text{bar}&=& (1.2 \pm 0.02) \Omega_0\\
    \Omega_\text{bar} &=& (1.4 \pm 0.04) \Omega_0\\
    \Omega_\text{bar} &=& (1.6 \pm 0.02) \Omega_0\\
    \Omega_\text{bar} &=& (1.74 \pm 0.01) \Omega_0
\end{eqnarray}
over which the OLR signature might be visible. We note again that selection effects might be obscuring part of the signature.

Secondly, the `$L_z$-range over which the OAS is visible' could be used as an additional constraint to identify the \emph{true} bar OLR. If we assume that the real MW bar is not much weaker than those in the simulation, the OLR signature \emph{should} in fact be visible over a range of $\Omega_\text{bar}$.
This rules out the candidate  $\Omega_\text{bar} = 1.74 \Omega_0$: Contrary to what we see in the simulations, the `green' stripe across angle space changes its slope from panel to panel around $\Omega_\text{bar} = 1.74 \Omega_0$; only in this single panel has it the correct slope of $-2$.

\subsection{The 1:1 resonance $(l=+1,m=1)$} \label{app:11_resonance}

For a bar perturbation with $m=2$, the second strongest resonance in the disk outside of CR after the OLR is the 1:1 outer Lindblad resonance. We showed an example 1:1 orbit in Figure \ref{fig:example_trapped_orbits}. The 1:1 resonance creates a high-$J_R$ scattering ridge analogous to the OLR, which can be seen, for example, in the upper left panel of Figure \ref{fig:angles_selection_effects_explained} at $L_z\sim1.4 L_{z,0}$.

If we select stars in our test particle simulations from this ridge, the expected $-m/l=-1$ slope in angle space is indeed revealed and could therefore be used to identify this resonance.

We searched for 1:1 signatures in the \emph{Gaia} angle space, but did not find any convincing candidates. This might be due to this resonance occurring at larger radius than the OLR and being therefore more difficult to detect. For realistic bar pattern speeds, the 1:1 resonance falls into a region in the Galactic disk that is (a) more strongly affected by selection effects and (b) less populated by stars. If the 1:1 falls at all into the survey volume, angle space $(\theta_\phi,\theta_R)$ will be less well sampled and appear more noisy.

As mentioned in \citet{2000AJ....119..800D} and in \citetalias{2021MNRAS.500.2645T}, there could be---at least locally for $d<200-600~\text{pc}$---a `red/blue' feature in action space around the 1:1 resonance for the \emph{fast} \texttt{Hercules} pattern speed. However, the corresponding angle space close to the \texttt{gold (H)} ridge in Figure \ref{fig:angle_space_scan_OLR} does not support this.

\subsection{The co-rotation resonance $(l=0)$} \label{app:CR_resonance}

The bar's CR strongly redistributes stars in $L_z$, which leaves characteristic patterns in the metallicity profile of the Galactic disk (e.g. \citealt{2021MNRAS.505.2412C,2021arXiv210505263W}). 

Based on our simulations, angle space, however, is not suited to identify CR. Within a survey volume around the Sun, no selection of stars based purely on action space was possible that isolated a high-enough fraction of resonant CR stars to make the $-m/l=-\infty$ (i.e. vertical) angle signature visible. This is because CR orbits oscillate around the bar's minor axis (Figure \ref{fig:example_trapped_orbits}) and not many are actually visiting the Solar neighbourhood (c.f., e.g., \citealt{2020ApJ...890..117D,2021arXiv210505263W}).

\subsection{The 1:4 resonance $(l=+1,m=4)$} \label{app:14_resonance}

Orbits at the 1:4 resonance follow $\Delta \theta_R/\Delta \theta_\phi = -4$ (see Figure \ref{fig:example_trapped_orbits}). However, the signatures that we expect in the cumulative stellar distribution around this resonance depend strongly on the choice of $m=4$ Fourier component of the bar model.

There are two aspects:
\begin{enumerate}[leftmargin=*,topsep=1ex,label=(\roman*)]
    \item \emph{Strength of the $m=4$ bar component.} To create a prominent high-$J_R$ overdensity ridge at the 1:4 resonance in action space, our test particle simulations required $m=4$ bar components with very strong $|\alpha_{m=4}|\gtrsim0.001$ (c.f. \citealt{2019AA...626A..41M}). This is consistent with the $m=4$ bar strength used in \citet{2018MNRAS.477.3945H} and  \citet{2019MNRAS.490.1026H}, $\alpha_{m=4} = 0.0015$, which is three times larger than the values quoted in these papers (J.A.S. Hunt, private communication).
    \item \emph{Orientation of the $m=4$ component.} \citet{2018MNRAS.477.3945H} used $\alpha_{m=4} < 0$, which aligns the potential minimum of the $m=4$ bar component with the potential maximum of the $m=2$ component, creating an overall boxy bar shape. They observed a `red/blue' feature, i.e. `Hercules/Horn'-like outward-/inward-motions, at the 1:4 resonance.\footnote{For $\alpha_{m=4} > 0$, which describes a pointy bar with ansae, \citet{2018MNRAS.477.3945H} observed a `blue/red' feature, i.e., the opposite way around as for the classic OLR or the boxy bar 1:4 feature.} \citet[their fig. 14]{2019MNRAS.490.1026H} explained this by two classes of 1:4 orbits: one that had pericenters close to the bar's major and minor axes (c.f. Figure \ref{fig:example_trapped_orbits}, $\theta_\text{slow}=0$), and one that was rotated by $45^\circ$ and had its apocenters along the bar's major and minor axes ($\theta_\text{slow}=\pi$). In our analogous simulations, only the former orbit class got populated and no 1:4 `red/blue' features were observed. Fig. 4 in \citet{2001AA...373..511F} suggests that these are stable orbits, while the $45^\circ$-rotated class of 1:4 orbits is unstable. It therefore depends on the kind of 1:4 orbits that get populated by stars, if and what kind of inward-/outward signature is observed at the 1:4 resonance.
\end{enumerate}

The orientation of the $m=4$ component and knowledge about the (re-)population history of different 1:4 orbit classes are essential for an analysis in $(L_z,J_R,v_R)$ space. For the analysis in $(\theta_\phi,\theta_R,\delta N)$, it might not matter that much, as this method is also sensitive to the symmetry of the orbit shape, and not just to the direction in which the stars move. 

In our test particle simulations with a boxy bar, we were only able to make the expected $\Delta \theta_R/\Delta \theta_\phi = -4$ signature visible in angle space, when we imposed a strong $m=4$ component with respect to the bar's basic quadrupole, e.g. $(\alpha_{m=2},\alpha_{m=4}) = (0.02,-0.0015)$. In this case, the trapping region and libration amplitude are large enough for a substantial ridge to develop at the 1:4 ARL in action space. The transition around the angle slope line for 1:4 orbits with $\theta_\text{slow}=\pi$ is from `green' to `red', similar to the $x_1(2)$ orbits in Figure \ref{fig:angle_space_model_OLR}.

In the case of a pointy bar and 1:4 orbits with $\theta_\text{slow}=0$, the expected transition is from `red' to `green', i.e. the opposite way around.

\section{The effect of using a different Galactic potential model} \label{app:Eilers_pot}

\begin{figure*}
    \centering
    \includegraphics[width=\textwidth]{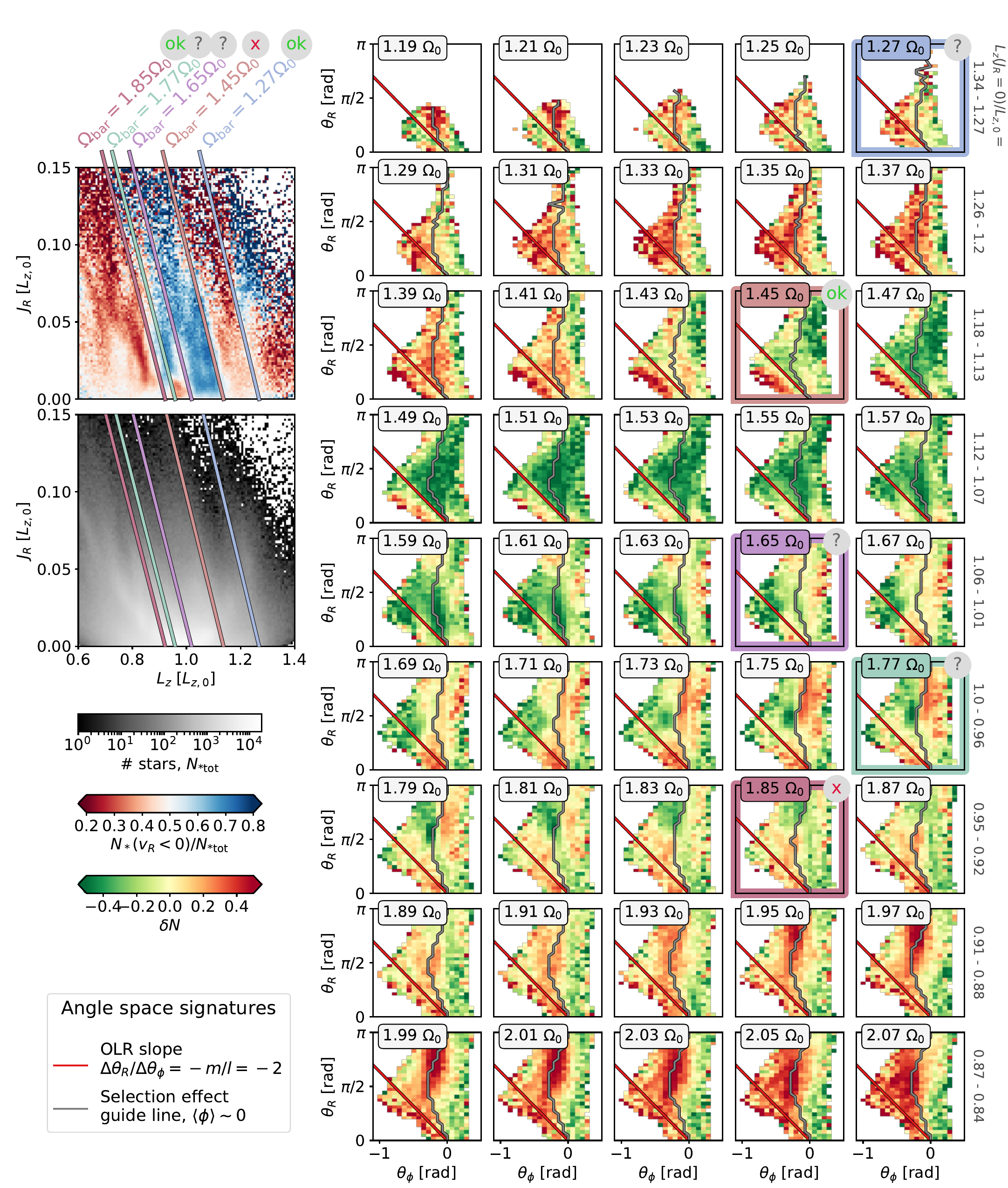}
    \caption{The \emph{Gaia} action-angle data for the MW potential model by \citet{2019ApJ...871..120E} (with $\Omega_0 = 28.3
~\text{km/s/kpc}$), analogous to Figures \ref{fig:angle_space_scan_OLR} and \ref{fig:angle_space_candidate_OLR}. OLR candidates derived from the two methods are compared: (i) slope of $-2$ in angle asymmetry on the \emph{right} vs. (ii) `red/blue' feature at the OLR ALR on the \emph{left}. The strongest candidate in angle space is also here close to the \texttt{Sirius} ridge, for $\Omega_\text{bar} = 1.45\Omega_0$, similar to the \emph{slightly faster S19B19} pattern speed.}
    \label{fig:angle_space_scan_OLR_Eilerspot}
\end{figure*}

Estimates of action-angle coordinates depend on the assumed gravitational potential model for the MW and the Sun's location in it. In the analyses of the \emph{Gaia} DR2 RVS data in this work, we have used the \texttt{MWPotential2014} model by \citet{2015ApJS..216...29B}, with the Sun at $R_0 = 8~\text{kpc}$, $v_\text{circ}(R_0) = 220~\text{km/s}$. Figure \ref{fig:angle_space_scan_OLR_Eilerspot} repeats this analysis for the MW potential model by \citet{2019ApJ...871..120E}, which uses $R_0 = 8.122~\text{kpc}$ \citep{2018AA...615L..15G}, and has $v_\text{circ}(R_0) = 229.8~\text{km/s}$. The pattern speed $\Omega_\text{bar}$ is in this case given in units of $\Omega_0 = v_\text{circ}(R_0) / R_0 = 28.3~\text{km/s/kpc}$. The general conclusions of this work remain unchanged when using this potential, but there are some subtle differences.

The OLR candidates identified in Section \ref{sec:OLR_candidates} are found at slightly different pattern speeds:
\begin{itemize}[leftmargin=*,topsep=1ex,itemsep=1ex]
\item $\Omega_\text{bar} = 1.27\Omega_0 = 35.9~\text{km/s/kpc}$, with the OLR at the \texttt{Hat}.
\item $\Omega_\text{bar} = 1.45\Omega_0 = 41.0~\text{km/s/kpc}$, with the OLR on the high-$L_z$ edge of the \texttt{Sirius} ridge.
\item $\Omega_\text{bar} = 1.65\Omega_0 = 46.7~\text{km/s/kpc}$, with the OLR between the `Hyades' and `Sirius'.
\item $\Omega_\text{bar} = 1.77\Omega_0 = 48.1~\text{km/s/kpc}$.
\end{itemize}

For the \citet{2019ApJ...871..120E} potential model, the angle plane at $\Omega_\text{bar} \sim 1.45\Omega_0$ (\emph{slightly faster slow bar}) looks even more convincing than in the \texttt{MWPotential2014} (Figure \ref{fig:angle_space_candidate_OLR}) when comparing to the model expectation for the OLR signature with angle slope -2. The angle plane for $\Omega_\text{bar} \sim 1.27\Omega_0$ (\emph{slow bar}), on the other hand, appears noisier; but because of the selection effects, we cannot rule this candidate out yet.

The pattern speed $1.77\Omega_0$ is also in the \citet{2019ApJ...871..120E} potential ruled out, because of the signature's variation with $L_z$ as argued in Section \ref{app:bar_strength_angle_space}.

The pattern speed of the \emph{fast bar}, $\Omega_\text{bar} \sim 1.85\Omega_0$, does also in the \citet{2019ApJ...871..120E} potential not exhibit the correct OLR slope and is therefore dismissed as a candidate, as discussed in Section \ref{sec:OLR_candidates}.

In Section \ref{sec:curious_coincidence} and in Figure \ref{fig:higher_order_resonances}, we found features in angle space with slopes of $-m/l$ at higher-order resonances $l=+1$ and $m\in\{3,4,5\}$. In this potential model, we reproduced this finding for $\Omega_\text{bar}=1.41\Omega_0=39.9~\text{km/s/kpc}$. The corresponding OLR in Figure \ref{fig:angle_space_scan_OLR_Eilerspot} is however not as convincing, as it misses the `green' part of the angle signature. The curious coincidence, that this pattern speed agreed also with the pattern speed of one of the OLR candidates, occurred therefore only for the \texttt{MWPotential2014}, but not for the \citet{2019ApJ...871..120E} potential.

%%%%%%%%%%%%%%%%%%%%%%%%%%%%%%%%%%%%%%%%%%%%%%%%%%

% Don't change these lines
\bsp	% typesetting comment
\label{lastpage}
\end{document}